\documentclass[acmsmall,screen,nonacm]{acmart}

\setcopyright{none}
\renewcommand\footnotetextcopyrightpermission[1]{}

\AtBeginDocument{%
}

\usepackage{hyperref}
\usepackage[hideedits,hidecomments]{collab}
\usepackage{enumitem}
\usepackage{balance}
\usepackage{xcolor}
\usepackage{subfigure}
\usepackage{multirow}
\usepackage{multicol}
\usepackage{tabularx}
\usepackage{siunitx}
\usepackage{float}
\usepackage{color}
\usepackage{diagbox}
\usepackage{caption}
\usepackage{makecell}
\usepackage{colortbl}
\usepackage{tcolorbox}
\usepackage{mdframed}
\usepackage{listings}
\usepackage{booktabs}
\usepackage[utf8]{inputenc}
\usepackage{calligra}
\usepackage[normalem]{ulem}
\usepackage{indentfirst}
\usepackage{url}
\usepackage{soul}
\usepackage{nicematrix}
\usepackage{fontawesome5}
\usepackage{textcomp}
\usepackage{stfloats}
\usepackage{verbatim}
\usepackage{graphicx}
\usepackage{amsmath,amsfonts}
\usepackage{algorithmic}
\usepackage{array}
\usepackage[utf8]{inputenc}
\usepackage{siunitx}

\definecolor{ballblue}{rgb}{0.13, 0.67, 0.8}
\definecolor{grey}{rgb}{0.9, 0.9, 0.9}
\definecolor{googlered}{rgb}{0.914, 0.262, 0.207}

\newcommand{\ie}{{\em i.e.},\xspace}
\newcommand{\eg}{{\em e.g.},\xspace}
\definecolor{mygray}{gray}{.9}

\newcommand{\boxmargin}{1mm}

\newtcolorbox{myboxa}[2][]{
    colback=gray!10!white,
    colframe=black, enhanced,
    attach boxed title to top left={yshift=-2mm,xshift=5mm},
    title=#2,#1
}
\newtcolorbox{myboxb}[2][]{
    boxsep=3pt,
    left = \boxmargin, right = \boxmargin, top = \boxmargin, bottom = \boxmargin,
    title={#2},#1
}
\newtcolorbox{myboxc}{
    colback=gray!15!white,
    arc = 0pt, outer arc = 0pt,
    boxsep=0pt, left = 3pt, right = 0pt, top = 0pt, bottom = 0pt, 
    leftrule=3pt, bottomrule=0pt,toprule=0pt, rightrule=0pt,
    left = \boxmargin, right = \boxmargin, top = \boxmargin, bottom = \boxmargin
}
\newtcolorbox{myboxd}{
    colback=gray!10,
    colframe=black,
    width=\columnwidth,
    arc=1mm, auto outer arc,
    boxrule=0.5pt,
}

\definecolor{myyellow}{HTML}{FFF2CC}
\newcounter{finding}

\begin{document}
\title[Lost in the Flow with Code Talkers]{Lost in the Flow with Code Talkers: Unveiling the Instruction-Tuning Tax of Large Language Models in Code Tasks}

\author{Shi Ying Chang}
\authornote{Both authors contributed equally to this research.}
\affiliation{%
  \institution{Singapore Management University}
  \city{Singapore}
  \country{Singapore}}
\email{sy.chang.2025@phdcs.smu.edu.sg}
\orcid{0009-0007-2606-8600}

\author{Chiok Yew Ho}
\authornotemark[1]
\affiliation{%
  \institution{The Chinese University of Hong Kong}
  \city{Hong Kong}
  \country{China}}
\email{cyho25@cse.cuhk.edu.hk}

\author{Yichen Li}
\affiliation{%
  \institution{The Chinese University of Hong Kong}
  \city{Hong Kong}
  \country{China}}
\email{ycli21@cse.cuhk.edu.hk}

\author{Yintong Huo}
\affiliation{%
  \institution{Singapore Management University}
  \city{Singapore}
  \country{Singapore}}
\email{ythuo@smu.edu.sg}

\renewcommand{\shortauthors}{Chang, Ho, et al.}

\begin{abstract}
AI coding assistants have significantly improved developer productivity by automatically suggesting code that aligns with user intent, and many of these tools are now integrated directly into Integrated Development Environments (IDEs). Developers interact with code in two distinct cognitive modes: \textit{Flow} and \textit{Command}. While developers require tools that directly complete or infill code in unfinished programs during Flow mode, they also need tools that can comprehend intentions expressed as natural-language instructions and convert them into executable code in Command mode. Although instruction-tuned Large Language Models (LLMs) dominate many application scenarios due to their abilities to infer and fulfill developers' intents, it remains unclear whether the same paradigm is equally suitable for different code-related tasks. Therefore, it is necessary to understand how instruction tuning affects the feasibility of CodeLLMs as coding assistants. To fill this gap, we conduct the \textbf{first empirical study that uncovers a key trade-off} caused by instruction tuning across programming modes, which we term the \textit{Instruction-Tuning Tax}. Our results show that instruction tuning is not a free lunch: although instruction-tuned models are more capable of following instructions and leveraging structured guidance, these gains often come at the cost of weaker infilling performance. We further extend our study through both qualitative and quantitative analyses, including manual failure categorization, behavioral metrics that capture generation fidelity, and intermediate-checkpoint evaluation throughout the tuning process. Summarizing our results into seven findings and four implications, our study offers a new perspective on the development of AI-powered coding tools and highlights the need to carefully balance instruction-following ability with effective code generation assistance.

\end{abstract}




\maketitle

\section{Introduction}
AI coding assistants have significantly improved developers' productivity by automatically suggesting code that aligns with their intentions.
Models such as \cite{copilot_doc,aiderAiderPair,cursorCursorCode} are no longer just research artifacts but are now integral components of the modern software development lifecycle, fundamentally altering how developers write, understand, and interact with code.

To maximize their utility and convenience, many are now integrated directly into Integrated Development Environments (IDEs). For instance, Aider \cite{aiderAiderPair} is a CLI-based AI coding assistant with an actively contributing community that is also integrated within IDEs as `Composer', and Cursor AI \cite{cursorCursorCode} is a similar coding editor that is able to understand working codebases and suggest code through surrounding code context or user instructions. 

\begin{figure*}[tbp]
    \centering
    \includegraphics[width=1.0\textwidth]{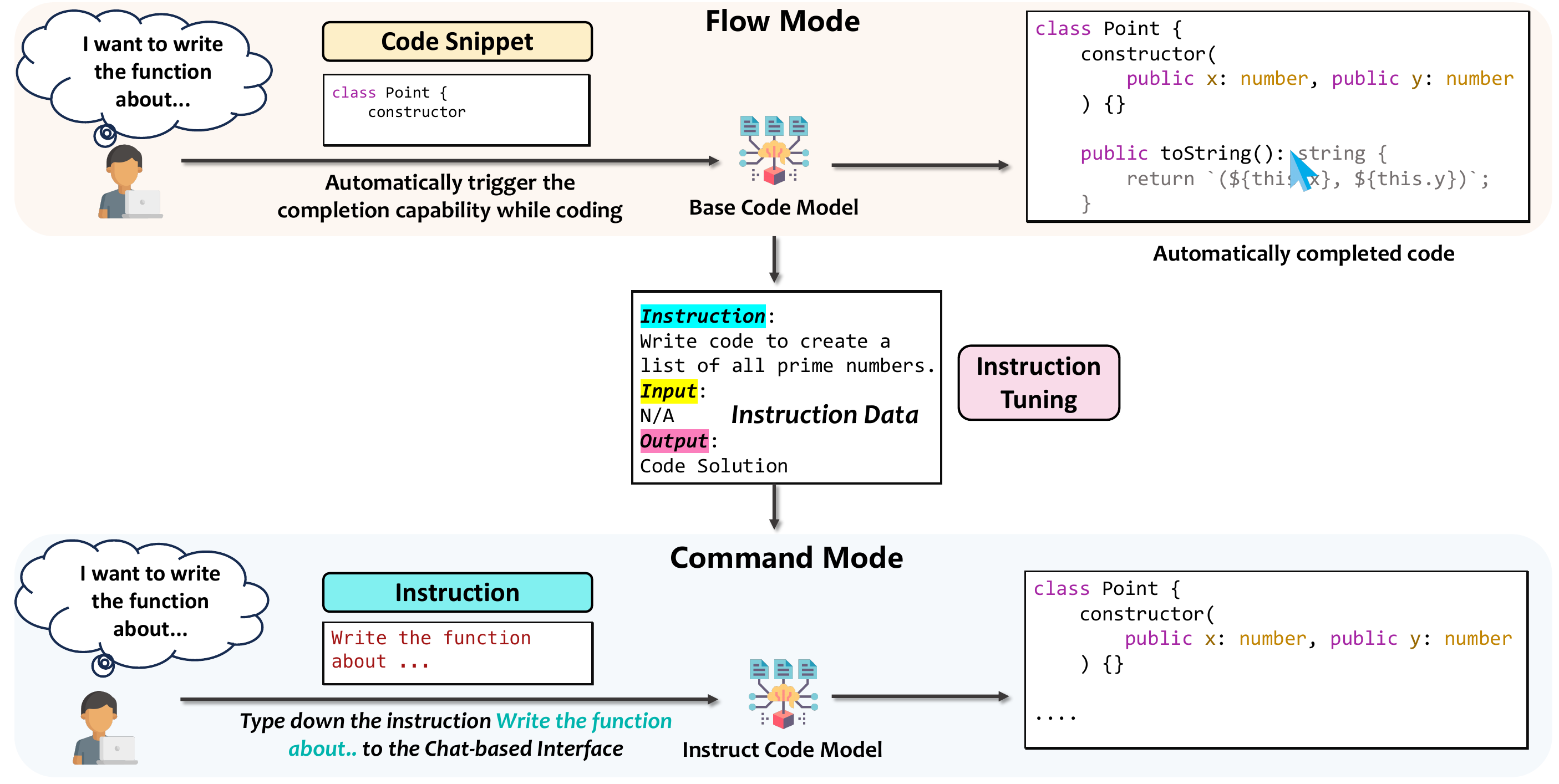}
    \vspace{-20pt}
    \caption{Flow mode vs. Command mode of modern coding tools.} 
    \label{fig: intro}
    \vspace{-10pt}
\end{figure*}

Yet, to exploit the potential of modern coding tools, we have to look beyond the models themselves and consider the developer's cognitive experience \cite{coding-assistant-hci}. According to current coding paradigms \cite{ai-assisted-prog-book, barke2023grounded}, a developer's interaction with code involves two distinct cognitive modes: \textbf{Flow} and \textbf{Command}. As shown in Figure~\ref{fig: intro}, in Flow mode, developers are fully immersed, thinking about the logic and syntax of code, with a focus on extending an existing thought process. In such cases, they need tools that directly suggest code lines to complete unfinished functionality, such as a missing logical block or an incomplete function. Specifically, the tool takes surrounding code contexts as inputs and suggests the next sequence of codes (i.e., \textit{completion}), or codes that fit into a missing ``hole'' within an incomplete code snippet (i.e., \textit{infilling}). In contrast, in Command mode, developers step back from the code to articulate the intents and specifications in natural language as \textit{instructions}, commonly when creating new functions or addressing new requirements. For this mode, tools must comprehend these instructions and convert human intents into executable code that developers can apply to their workflow. 

The use of Large Language Models (LLMs) has become prevalent within our daily lives, where most of them release two distinct variants: \textbf{base} model and \textbf{instruct} model, as illustrated in Figure~\ref{fig: intro}. While base models are pre-trained to perform text completion based on the given input \cite{llama2}, instruct models are further fine-tuned with instruction-response pairs and objectives that guide them to better align with human preferences and values \cite{ouyang22training}. Code language models (so-called CodeLLMs) have similar training strategies by continually pre-training and tuning on code-related instruction-response pairs for the instruct variant, such as Qwen2.5-Coder and DeepSeek-Coder \cite{deepseekcoder,qwen25coder}. 
Instruct models like ChatGPT \cite{chatgpt}, dominate most of the use cases because they can naturally and conversationally respond to a user's direct requests.

The distinct cognitive modes in programming, flow and command, necessitate different model capabilities and contextual suggestions. Intuitively, a model's pre-training on infilling objectives seems better aligned with flow mode tasks like code completion, whereas instruction-tuning simulates interactions in command mode. Therefore, we pose the question: \textit{Would the tuned ones inherit advantages both from pre-training and instruction-tuning phase, yielding a unified and ideal solution for programming?} 
However, several hurdles in this continuous tuning process can degrade model performance: (i) the explicit special tokens (i.e., prefix, middle/hole, and suffix) required for the pre-training infilling objective may not generalize effectively during the tuning stage \cite{infilling-bavarian};
(ii) subsequent training risks overwriting pre-learned abilities \cite{mixing-composition}, known as catastrophic forgetting (CF) \cite{cf-llm}. 
Therefore, we need to better understand if the abilities of current CodeLLMs actually align with what programmers need. A lack of this understanding will impede the development of practical and effective AI-powered coding tools. 

To fill in the gap, we conduct the first study that reveals a key \textit{trade-off} caused by instruction tuning across programming modes, which we term the \textit{instruction-tuning tax}. Concretely, we empirically evaluate how instruction-tuned CodeLLMs perform on code completion, infilling, and generation tasks after being fine-tuned to follow instructions. Our results show that \textit{instruction tuning is not a free lunch}: teaching a base model to follow coding instructions strengthens its command-mode ability, but often comes at the cost of weaker infilling performance and shifts in generation behavior, as reflected by both execution-based correctness (Pass@1) and behavioral metrics.

More specifically, we formulate the following research questions:
\textit{(RQ1)} What is the performance difference between base and instruct models on representative coding tasks?  
\textit{(RQ2)} What are the underlying reasons for the performance differences implied by the change in models' behavior? 
\textit{(RQ3)} How does model performance evolve during the fine-tuning process from Base to Instruct?

To answer \textbf{RQ1}, we identified representative benchmarks for each of the coding tasks to expose and quantify the instruction-tuning tax of open-source CodeLLMs. 

We also extend this to models of different parameter sizes under the same model family to investigate the effect of instruction-tuning tax with respect to model sizes. For \textbf{RQ2}, we manually sampled generation results from the evaluated models and examined the failure cases that reflected different aspects of model behavior. Based on these observations, we derived behavioral metrics to quantify recurring patterns in the generated outputs and assess their practical usability in Flow mode. For \textbf{RQ3}, we adopted the open-source CodeLLM Magicoder \cite{magiccoder} to fine-tune Qwen2.5-Coder-7B-Base as a representative mid-sized model. The resulting intermediate models were then evaluated using the settings from \textbf{RQ1}, allowing us to examine how model behavior changes across fine-tuning stages.

Our study reveals that the instruction-tuning tax is not uniform across model families. In particular, Qwen2.5-Coder instruct models are generally more robust than DeepSeek-Coder on infilling and completion tasks, although both families exhibit clear trade-offs after instruction tuning. While instruct models tend to perform better when tasks provide structured natural-language guidance, base models still retain advantages on fill-in-the-middle tasks, where they solve more exclusive problems and exhibit different failure patterns from their instruction-tuned counterparts. Furthermore, intermediate-checkpoint analysis shows that instruction tuning induces non-monotonic but systematic capability shifts: instruction-following abilities improve, infilling performance generally declines, and completion exhibits benchmark-dependent behavior, with aggregate gains that weaken in later tuning stages. Based on these findings, we derive implications for both researchers and practitioners toward the better development and deployment of AI coding assistants in the future. Specifically, our contributions are threefold.

\begin{itemize}
    \item We conduct the first empirical study to uncover the \textit{Instruction-Tuning Tax}, a key trade-off introduced by instruction tuning across Flow-based and Command-based coding tasks.
    \item We extend the study through qualitative and quantitative analysis of model behavior, including manual failure analysis and behavioral metrics that capture generation fidelity.
    \item We analyze intermediate checkpoints to trace performance shifts throughout the instruction-tuning process.
    \item We derive two implications for researchers and two for practitioners, aiming to inform the future development and deployment of AI coding assistants.
\end{itemize}

\section{Background and Challenges}

The paradigm of automated code generation has undergone a significant transformation with the advent of LLMs. This section outlines the evolution of code-related models, the role of instruction tuning, and three challenges in investigating the trade-off in the tuning process. 

\begin{figure*}[tbp]
    \centering
    \includegraphics[width=1.0\textwidth]{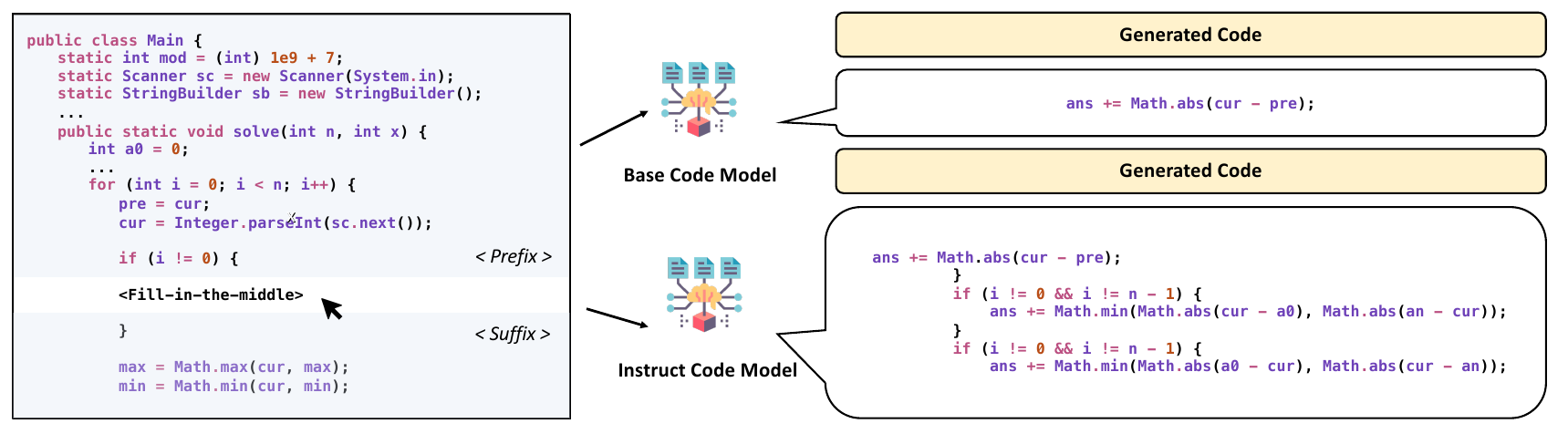}
    \vspace{-20pt}
    \caption{A motivation example on the difference between the base model and the instruct model.} 
    \label{fig: motivation}
    \vspace{-10pt}
\end{figure*}

\subsection{The Pretraining of Code LLMs}

Early work on code generation mainly relied on encoder-decoder architectures, such as CodeT5~\cite{codet5} and CodeT5+ ~\cite{codet5plus}, which showed strong capabilities on code synthesis. Under this architecture, the encoder processes input sequences of code or text to learn contextual representations, while the decoder is subsequently trained to output code based on these representations.
More recently, the field has changed to utilize decoder-only transformer models. 
Such an approach, first adopted by Codex~\cite{codex} and later scaled in models like StarCoder \cite{starcoder}, DeepSeek-Coder \cite{deepseekcoder}, and Qwen-Coder \cite{qwen25coder}, is built on auto-regressive pretraining of vast code corpora.

The core objective of auto-regressive pretraining is next-token prediction: given a sequence of preceding tokens, the model is trained to predict the next token. This objective teaches the model statistical regularities of programming languages, including syntax, structure, and common coding patterns~\cite{brown2020language,chen2021evaluating}. 
For code generation, two widely used context modeling strategies are unidirectional left-to-right modeling and fill-in-the-middle (FIM) prediction~\cite{guo2022unixcoder,incoder,infilling-bavarian}. 
The former supports standard code completion by generating code conditioned on preceding context, while the latter trains the model to recover a missing code span given both prefix and suffix context. 
These strategies can also be combined, allowing CodeLLMs to benefit from both left-to-right continuation and bidirectional contextual constraints.
These capabilities, which we term structure-aware generation, are fundamental to a model's utility within a developer's workflow and represent the direct outcome of the resource-intensive pre-training process.

\subsection{From Base Code Models to Instruction Code Models}

Although pre-training empowers code LLMs with comprehensive, statistical knowledge of code, it fails to align with human preferences. 
To bridge this gap between latent capability and practical application, researchers introduce instruction tuning. This supervised fine-tuning (SFT) process trains the base model on a dataset of instruction-response pairs, aligning natural language commands~\cite{ouyang2022training} (\eg \textit{Write a function that sorts a list of strings by length}) with their corresponding code implementations.

Figure~\ref{fig: intro} illustrates the tuning process from base to instruct variants (\eg Qwen2.5-Coder and Qwen2.5-Coder-Instruct), and their corresponding common application scenarios. 
More specifically, the instruct variant is created by applying supervised fine-tuning (SFT) to the base model using a dataset of instruction-response pairs. This subsequent training phase endows the model with the ability to interpret natural language and map it to specific code outputs, thereby aligning its generative capabilities with user intent.

The success of instruction tuning comes from high-quality, large-scale instruction datasets, and methods for building those datasets have evolved much. Early work, such as Code Alpaca~\cite{alpaca}, applies the \textit{Self-Instruct} technique~\cite{wang2022self}, which enables stronger LLMs to take a small set of human-written seed prompts and expand them into a diverse, synthetic corpus of instruction–response examples. Later efforts focused on increasing the difficulty and diversity of synthetic data. For example, the Evol-Instruct procedure used by WizardCoder~\cite{wizardcoder} iteratively rewrites simple instructions into more complex variants to expose models to a wider range of problem-solving scenarios. At the same time, researchers have tried to reduce the reliance on powerful teacher models. The \textit{OSS-Instruct} approach, used by Magicoder~\cite{magiccoder} and StarCoder2-instruct~\cite{starcoder2}, feeds real open-source code snippets to a model and asks it to generate the corresponding instruction, producing the high-quality instruction data.

\subsection{Challenges in revealing the trade-off}

One motivating example is shown in Fig.~\ref{fig: motivation}, the instruction-tuned model outputs an additional conditional branches, causing execution errors and illustrating the common tendency of such models to generate unnecessary surrounding code.
To systematically evaluate and understand the side effects of instruction tuning for different programming modes, we need to overcome the following challenges.

\textbf{1) Select representative problems.} 
A broad set of programming-related tasks is being proposed these days, such as code generation, program repair, and automatic testing. Because these tasks address different stages and scenarios of the software lifecycle, it is important to choose the tasks that best match our evaluation goals.

\textit{Solution.} Considering our aim to expose effects in two cognitive modes, flow and command, we then abstract these two modes into three representative programming tasks: code completion (flow mode), code infilling (flow mode), and code generation (command mode). 
To ensure diversity and representativeness, selected problems should come from multiple sources (GitHub, StackOverflow, and programming platforms), be constructed by multiple strategies (hand-written problems, automated construction, and crawled examples from developer forums that reflect real-world practice), and cover multiple programming genres such as API handling (e.g., use of Python packages), class declarations, conditionals and logical operations in algorithmic problems.

\textbf{2) Reveal behavioral shifts in model-generated code.}
Even when benchmarks apply a unified post-processing procedure to all models, they do not necessarily reveal the behavioral shifts introduced by instruction tuning. Most self-contained code generation benchmarks adopt rule-based post-processing on a best-effort basis to normalize outputs before execution, which helps ensure fairness in benchmarking. However, this also circumvents a more fundamental question: what kinds of output deviations are models inherently prone to produce, and how do these tendencies differ across model variants? In particular, unified post-processing does not capture whether some models are more likely to generate redundant code, irrelevant natural language, special tokens, or structurally unnecessary continuations. These behavioral biases are important beyond benchmark evaluation, as they directly affect real-world deployment settings such as IDE completion and infilling, where users expect concise and well-bounded code suggestions rather than outputs that merely become executable after benchmark-specific cleanup.

\textit{Solution.} We therefore go beyond standard benchmark-level correctness and design methodologies to explicitly analyze behavioral attributes in model-generated code. In addition to comparing functional outcomes, we manually inspect failure cases and categorize recurrent deviations introduced by instruction tuning. These observations further motivate the design of auxiliary metrics that quantify generation discipline, enabling us to study not only whether a model solves the task, but also how it behaves while doing so. Such analysis is necessary for understanding the practical impact of instruction tuning and for informing more general post-processing and deployment strategies in real-world coding assistants.

\textbf{3) Demystify the black-box tuning process.}
While unified quantitative evaluations can precisely measure performance shifts following instruction tuning, they treat the base and instruction-tuned models as black boxes. This approach fails to elucidate why performance varies and how tuning fundamentally alters a model's intrinsic capabilities.

\textit{Solution.} To demystify the SFT process, we first perform a comparative analysis of outputs from base and instruction-tuned models, concentrating on cases where the models produce incorrect answers. This comparison isolates the direct effects of instruction tuning. 
Besides, we replicate the fine-tuning pipeline and divide it into discrete stages. By evaluating intermediate checkpoints at each stage, we trace how varying degrees of instruction tuning alter performance across different programming tasks.

\section{Study Design}

\begin{figure*}[tbp]
    \centering
    \includegraphics[width=1.0\textwidth]{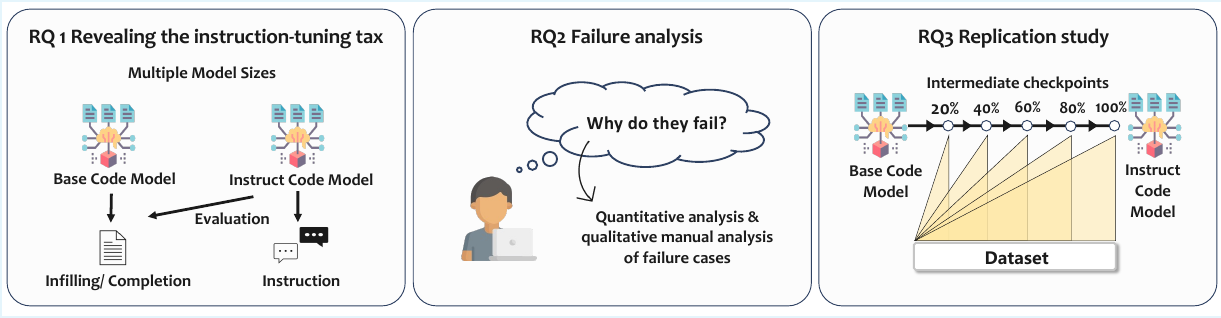}
    \vspace{-20pt}
    \caption{Experimental settings of our study design corresponding to the research questions.}
    \label{fig: experimentalsettings}
    \vspace{-22pt}
\end{figure*}

Our overall study design is presented in Figure \ref{fig: experimentalsettings}.
This paper quantifies the cost of instruction tuning on diverse coding tasks by evaluating top-performing CodeLLMs and their variants. Our study further analyzes and discusses performance differences across model variants and sizes, as well as to track performance evolution throughout the tuning process.  In addition to quantitative results, we present quantitative and qualitative analyses that explain the underlying causes of task failures and motivate the design of various behavioral metrics for further evaluations.

Specifically, we start with exposing the existence of the instruction-tuning tax through RQ1. That is, CodeLLMs exhibit performance degradation on infilling tasks that are learned during pretraining, and we also consider multiple model sizes to evaluate whether larger parameter counts mitigate the trade-off introduced by instruction tuning as well as models across families (Qwen2.5-Coder and DeepSeek-Coder) to observe model-wise effects. 
RQ2 performs qualitative analyses to identify failure causes for performance degradation in RQ1 and informs the design of behavioral metrics to measure the fidelity of model-generated codes.
As for RQ3, we replicate the instruction-tuning process of Magicoder, an open-sourced and open-data CodeLLM, and collect intermediate checkpoints to trace how the model evolves through different tuning phases.

\subsection{Task formulation}

\begin{figure*}[tbp]
    \centering
    \includegraphics[width=1.0\textwidth]{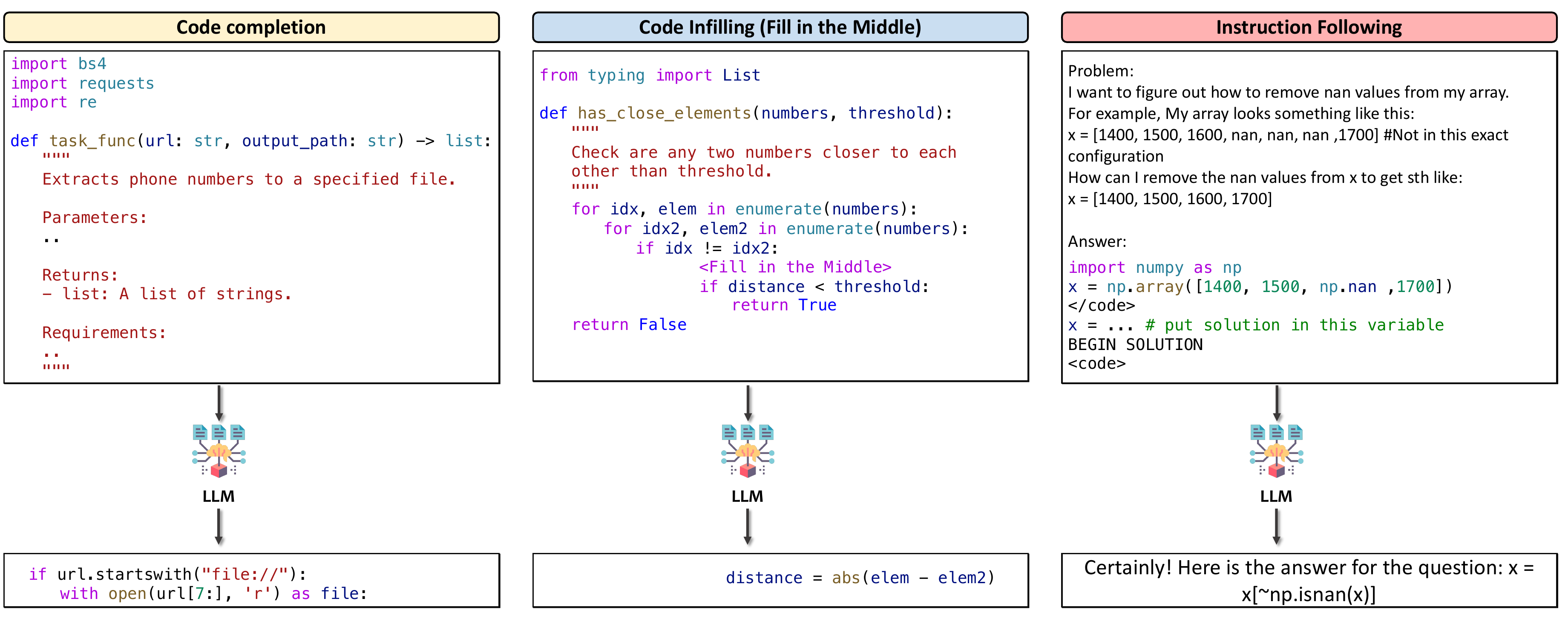}
    \vspace{-20pt}
    \caption{Task Formulation with Examples.}
    \label{fig: taskformulation}
    \vspace{-10pt}
\end{figure*}

We discuss the task formulation and conceptualization involved in our experiments and illustrate them in Figure~\ref{fig: taskformulation} . The notation `||' used thereafter represents concatenation. We elaborate our benchmark implementations in details in the upcoming section.

\subsubsection{Code Infilling (Fill-In-the-Middle, FIM)}
Code infilling, or Fill-in-the-Middle (FIM), is a code completion paradigm in which a model must predict a missing fragment (i.e., holes) inside a code snippet. Formally, an arbitrary code snippet $C = \{p||m||s\}$ can be split into three parts: prefix ($p$), middle ($m$) and suffix ($s$), where model’s objective is to generate $m$ given $p$ and $s$.
When preparing FIM inputs for LLMs, the typical format (i.e., FIM template) uses explicit markers to identify each segment. One common template is: ``\texttt{<fim\_prefix>} $p$ \texttt{<fim\_suffix>} $s$ \texttt{<fim\_middle>} $m$'', where special tokens identify the prefix, suffix, and middle \cite{infilling-bavarian}.

In the case of our infilling benchmark tasks, guidance are provided in the form of inline code comments or docstrings to inform the models of the intended functionality.
For our experiments that involve this task, the FIM template shown above is adopted to take in two elements, prefix ($p$) and suffix ($s$) as input, enabling the model to output the infilling target, $m$ that can be appended after the middle token.

\subsubsection{Instruction Following}
The instruction-following paradigm encodes the task description and any supporting context into a single instruction prompt. Models are typically given inputs using a chat template that includes special tokens to mark the instruction and the expected response. A common template (i.e., chat template) is: ``\texttt{<SYSTEM>} [system prompt] \texttt{<INST>} [instruction] \texttt{<RESPONSE>} [response]'' where \texttt{<INST>} and \texttt{<RESPONSE>} are special tokens that signal the model where the instruction ends and the response should begin, and the system prompt is used to define the role of the model or dictate specific requirements that the model should obey. This task limits to instruct models. For tasks that involve instructions, we take the context of the problem (i.e., instructions and relevant code contexts) as [instruction] and apply chat templates for the respective instruct models. Additionally, we also define system prompts for tasks that require strict output formats. As such, the models respond in the anticipated format for evaluation to address the problems given in contexts as instructions.

\subsubsection{Code Completion}
Code completion requires a model to complete a code snippet by generating subsequent codes that succeed the existing code. It requires the model to suggest codes directly in a left-to-right manner, which follows the conventional next token prediction paradigm \cite{gpt3}. In our work, we treat completion as a \textit{guided} task: the instruction and any necessary context are embedded in comments or docstrings, and the model must produce a coherent continuation of the code when provided with sufficient contextual information to generate reasonable code. The problems of guided completion comprise a function declaration followed by guidance wrapped in docstrings. As such, the models complete the functions by outputting the subsequent codes that realize the target functionalities.

\subsection{Experimental Settings}
With reference to our study design illustrated in Figure \ref{fig: experimentalsettings}, we discuss the experimental settings and implementations in detail below.

\subsubsection{Models}
We select the Qwen2.5-Coder family~\cite{qwen25coder} and the DeepSeek-Coder family~\cite{deepseekcoder} as the CodeLLMs studied in this work for two reasons. First, they provide broadly comparable model scales across Base and Instruct variants, spanning small, medium, and large parameter ranges. Second, both families are among the strongest open-source code models on public benchmark leaderboards~\cite{BigCodeModelsLeaderboard} and are widely studied in both academic and industrial settings~\cite{wang2025rag}. For RQ1, we evaluate Qwen2.5-Coder models with 1.5B, 7B, 14B, and 32B parameters, and DeepSeek-Coder models with 1.3B, 6.7B, and 33B parameters. For RQ3, we use Qwen2.5-Coder-7B as a representative mid-sized code model, as it offers a practical balance between capability and tuning cost while allowing us to replicate the Magicoder pipeline and collect intermediate checkpoints in a controlled and reproducible setting.

\subsubsection{Benchmarks}

\begin{table}[t]
\centering
\footnotesize
\setlength{\tabcolsep}{4pt}
\caption{Statistics of the benchmark datasets used in this study. RQ1 and RQ2 evaluate Fill-in-the-Middle and Completion benchmarks only, while RQ3 additionally includes Instruction Following benchmarks.}
\label{tab:dataset}

\newcolumntype{Y}{>{\raggedright\arraybackslash}X}
\begin{tabularx}{\linewidth}{@{}p{2.7cm}|Y Y p{2.5cm} r@{}}
\toprule
\textbf{Task} & \textbf{Dataset} & \textbf{Problem Source} & \textbf{Language} & \textbf{Size} \\
\midrule
\multirow{3}{*}{Fill-in-the-Middle}
& HumanEval Infilling & Human-Authored & Python & 1,033 \\
& SAFIM (API, Block, Control) & Codeforces and GitHub & Python, C++, Java, C\# & 17,720 \\
& ClassEval-LineInfilling & Human-Authored (ClassEval) & Python & 2,557 \\
\midrule
\multirow{2}{*}{Completion}
& BigCodeBench-Complete & ODEX seed examples & Python & 1,140 \\
& ClassEval-Completion & Human-Authored (ClassEval) & Python & 396 \\
\midrule
\multirow{4}{*}{Instruction Following}
& DS-1000 & StackOverflow & Python & 1,000 \\
& BigCodeBench-Instruct & ODEX seed examples & Python & 1,140 \\
& HumanEval (HumanEval(+)) & Human-Authored & Python & 164 \\
& MBPP (MBPP(+)) & Crowd-sourced & Python & 399 \\
\bottomrule
\end{tabularx}

\end{table}
We organize our evaluations around three benchmark categories: \emph{Fill-in-the-Middle (FIM)}, \emph{Completion}, and \emph{Instruction Following}. \textbf{RQ1} and \textbf{RQ2} evaluate only FIM and Completion, since these settings allow a more controlled comparison between base and instruction-tuned code models without introducing differences in instruction-format alignment. \textbf{RQ3} additionally includes Instruction Following benchmarks to study how capabilities evolve during instruction tuning.

For the FIM task, we use HumanEval Infilling (HEFIM) \cite{infilling-bavarian}, SAFIM \cite{safim} and ClassEval-LineInfilling covering single-line and block-wise infilling scenarios. 

\begin{itemize}
\item \textbf{HumanEval Infilling} (HEFIM) is a variant of the HumanEval benchmark \cite{humaneval}, which is transformed by removing lines from the problem solutions as infilling questions. We implemented the single-line infilling benchmark task with the prefix and suffix provided by HumanEval Infilling (HEFIM), the former being labelled as `prompt' in the original data column. That is, we apply the infilling template by directly taking the prefix and suffix for each problem in the HEFIM dataset.
\item \textbf{SAFIM} is a benchmark dataset consisting of syntax-aware, multi-line completion problems that involve completing conditions and code blocks in multiple programming languages. There are 3 subsets of SAFIM: API, Block and Control, with API focusing on the use of common API packages and the latter two on completing logical code blocks and conditions in algorithmic problems. As the original dataset problems are given with a hole in the code replaced with the ``\{\{completion\}\}'' tag, we split the problem codes by this tag and obtain the prefix and suffix for each problem that can be applied to the infilling template.
\item \textbf{ClassEval-LineInfilling} (ClassEval-Infilling) is derived from the ClassEval \cite{classeval} dataset, a human-written benchmark consisting of 100 class-level coding instances. To produce infilling instances, each non-empty line in the body of every non-constructor method is converted into an infilling instance consisting of a full-file prefix, target line, and suffix. 
\end{itemize}

For the Completion task, we adopt BigCodeBench-Complete and ClassEval-Completion.

\begin{itemize}

\item \textbf{BigCodeBench-Complete} is the completion variant of BigCodeBench, which contains 1,140 programming tasks assessing advanced coding capabilities such as diverse function usage, tool interaction, and multi-step reasoning.

\item \textbf{ClassEval-Completion}  is derived similar to its infilling counterpart: each non-constructor method is removed from the full class implementation and replaced with its corresponding skeleton specifying the required context of the codes to be implemented (i.e., to-dos), yielding a method-level completion input. Both derivations preserve the original class-level semantic context while changing the prediction granularity from whole-class generation to line-level infilling or method-level completion.

\end{itemize}

For RQ3, we additionally include Instruction Following benchmarks, as this part of the study examines how model capabilities evolve throughout instruction tuning rather than only comparing the final base and instruction-tuned endpoints. Beyond the FIM and Completion benchmarks used in RQ1 and RQ2, RQ3 therefore incorporates four instruction-following datasets: DS-1000 \cite{ds1000}, BigCodeBench-Instruct \cite{zhuo2024bigcodebench}, HumanEval, and MBPP. Together, these benchmarks cover a broad spectrum of programming scenarios, including data science workflows, functionally complex coding tasks, and canonical instruction-based code generation problems.

\begin{itemize}
\item \textbf{DS-1000} consists of 1,000 programming problems centered on data science applications across seven widely used Python libraries. The tasks are curated from StackOverflow and reflect realistic instruction-style coding problems grounded in practical library usage. We use the dataset's \texttt{prompt} field as the task instruction. To encourage the model to generate executable code snippets rather than free-form explanations or full scripts, we prepend three fixed in-context demonstrations before each prompt. We further apply a system prompt stating: ``You will be given Prompt: (It may include Problem: and A: in the prompt), and you must output only the codes to be continued after the BEGIN SOLUTION or SOLUTION START.'' This helps align the model with the expected output format and reduces verbose responses.

\item \textbf{BigCodeBench-Instruct} is the instruction-following variant of BigCodeBench, containing 1,140 programming tasks that assess advanced coding abilities beyond conventional benchmark suites. The tasks involve realistic function usage, diverse library and tool interactions, and multi-step reasoning, making this benchmark useful for analyzing how later stages of instruction tuning affect instruction-conditioned coding behavior.

\item \textbf{HumanEval} \cite{humaneval} and \textbf{MBPP} \cite{mbpp} are two widely used benchmarks for instruction-based code generation in Python. HumanEval evaluates functional correctness on human-written programming tasks using unit tests, while MBPP consists of crowd-sourced Python problems expressed in natural language. In our evaluation, we follow the EvalPlus-based protocol \cite{evalplus} adopted in the Magicoder experiments \cite{magiccoder}, and therefore report results as \emph{HumanEval (HumanEval(+))} and \emph{MBPP (MBPP(+))}. Here, the main score corresponds to the original benchmark, while the value in parentheses corresponds to the stricter EvalPlus version with expanded test coverage. This enhanced evaluation allows us to preserve comparability with prior code-generation results while also reflecting robustness under more rigorous evaluation.
\end{itemize}

\newcolumntype{Y}{>{\raggedright\arraybackslash}X}

\subsubsection{Metrics}
For the mentioned benchmarks, we use the official evaluation pipelines, executed locally using Pass@1 to assess model performance. Exact match (EM) is used for ClassEval-Infilling which is single-line infilling tasks that require the output line to match exactly as the reference line string. For MBPP(+) and HumanEval(+), we follow the Magicoder pipeline, also reporting Pass@1. To ensure consistent evaluation, we normalized outputs by removing extraneous whitespace through stripping, which was particularly important for the DeepSeek-Coder model. Despite performing output post-processing with best efforts to rule out formatting mismatches such as indentations, we are unable to guarantee coverage of all erroneous output formats. To this end, we derive additional metrics such as Code-to-Token Ratio (CTR) and the presence of natural language markers and extra classes/functions for a more holistic evaluation on the fidelity of generated codes, which will be addressed and discussed further in RQ2.

\subsubsection{Instruction Tuning} 
To investigate the effect of instruction-tuning on models' performance in a progressive manner (RQ3), we adopted the Magicoder \cite{magiccoder} pipeline because it released scripts and datasets for replication. We follow the same two-step sequential tuning process in line with the Magicoder implementation: the first phase uses Magicoder-OSS-Instruct-75K dataset, then continued tuning with Magicoder-Evol-Instruct-110K as the second phase. 

To obtain intermediate models along the tuning process, we sampled different fractions of the training data at each phase of the Magicoder pipeline. For the OSS-Instruct-75K stage, we created progressively larger training subsets using 5\%, 10\%, 20\%, 40\%, 60\%, 80\%, and 100\% of the data. For the Evol-Instruct-110K stage, we used 20\%, 40\%, 60\%, 80\%, and 100\% of the data. This yields 12 checkpoints in total across the two stages. Notably, all checkpoints in the Evol-Instruct phase are obtained by further tuning the model that has already been trained on the full OSS-Instruct-75K dataset. We also compared our reproduced full model with Magicoder’s officially released model to verify the fidelity of replication, and then analyzed long-run performance drift by comparing checkpoint performance against the initial base model.

\subsubsection{Time and GPU Utilization}
All experiments are conducted on NVIDIA A100 80GB GPUs, except for large-parameter models such as DeepSeek-Coder-33B, which require the H200 140GB GPU to meet memory constraints. Output generation time varies substantially across benchmarks. For smaller benchmarks, full generation on HumanEval (HumanEval(+)) takes roughly 10 minutes per model, while MBPP (MBPP(+)) requires about 17 minutes. SAFIM-API typically takes 10--30 minutes, and ClassEval-Completion generally requires more than 30 minutes. HumanEval Infilling (HEFIM) completes in about 1 hour, while ClassEval-LineInfilling and DS-1000 typically require 1--2 hours depending on model size. The SAFIM Block and Control variants roughly took 4--6 hours. BigCodeBench is the most computationally intensive benchmark; for example, generating outputs for Qwen2.5-Coder-7B-Instruct requires around eight hours. Across all tasks, we consistently observe that instruction-tuned models take longer to evaluate than their base counterparts. Fine-tuning costs also scale with data size. On Qwen2.5-Coder-7B, fine-tuning 5\% of the 75K dataset for two epochs requires approximately 20 minutes, while 10\% doubles to about 40 minutes. These observations highlight computational trade-offs: instruction-tuned models incur longer inference times, and fine-tuning runtime grows nearly linearly with data size.

\section{Study Results}

\subsection{RQ1: What is the performance difference between Base and Instruct models on representative coding tasks?}

To answer this as our fundamental RQ, models from the Qwen2.5-Coder family \cite{qwen25coder} and the DeepSeek-Coder family \cite{deepseekcoder} are evaluated against representative benchmarks of each tasks. The results are reported in Table \ref{tab:rq1-qwen} and \ref{tab:rq1-dsc}.

\begin{table}[t]
\centering
\scriptsize
\setlength{\tabcolsep}{3.5pt}
\caption{Evaluation results of Qwen2.5-Coder models on Fill-in-the-Middle and Completion tasks. Brackets for instruct models indicate the relative difference compared to their corresponding base models.}
\label{tab:rq1-qwen}
\resizebox{\columnwidth}{!}{%
\begin{tabular}{llllllll}
\toprule
\textbf{Qwen2.5-Coder} &
\multicolumn{3}{c}{\textbf{Fill in the Middle}} &
\multicolumn{2}{c}{\textbf{Completion}} \\
\cmidrule(lr){2-4} \cmidrule(lr){5-6}
& \textbf{HEFIM} & \textbf{SAFIM} & \textbf{ClassEval-Infilling} &
\textbf{BigCodeBench} & \textbf{ClassEval-Completion} \\
\midrule
1.5b-base & 43.37 & 39.61 & 74.93 & 26.20 & 37.87 \\
7b-base   & 70.95 & 47.95 & 79.94 & 39.74 & 35.85 \\
14b-base  & 73.28 & 62.89 & 82.48 & 51.80 & 36.11 \\
32b-base  & 72.99 & 69.70 & 84.16 & 54.60 & 54.79 \\
\midrule
1.5b-instruct & 61.94\,(+42.24\%) & 40.23\,(+1.57\%) & 67.38\,(-10.08\%) & 29.60\,(+12.98\%) & 48.73\,(+28.68\%) \\
7b-instruct   & 61.86\,(-12.81\%) & 46.35\,(-3.33\%)  & 74.93\,(-6.29\%)  & 46.50\,(+17.01\%) & 58.33\,(+62.71\%) \\
14b-instruct  & 54.89\,(-25.10\%) & 55.31\,(-12.06\%)  & 78.84\,(-4.41\%)  & 57.50\,(+11.00\%) & 63.63\,(+76.21\%) \\
32b-instruct  & 70.28\,(-3.71\%)  & 57.38\,(-17.67\%)  & 80.09\,(-4.84\%)  & 59.90\,(+9.71\%)  & 59.59\,(+8.76\%) \\
\bottomrule
\end{tabular}
}
\end{table}

\begin{table}[t]
\centering
\scriptsize
\setlength{\tabcolsep}{3.5pt}
\caption{Evaluation results of DeepSeek-Coder models on Fill-in-the-Middle and Completion tasks. Brackets for instruct models indicate the relative difference compared to their corresponding base models.}
\label{tab:rq1-dsc}
\resizebox{\columnwidth}{!}{%
\begin{tabular}{llllllll}
\toprule
\textbf{DeepSeek-Coder} &
\multicolumn{3}{c}{\textbf{Fill in the Middle}} &
\multicolumn{2}{c}{\textbf{Completion}} \\
\cmidrule(lr){2-4} \cmidrule(lr){5-6}
& \textbf{HEFIM} & \textbf{SAFIM} & \textbf{ClassEval-Infilling} &
\textbf{BigCodeBench} & \textbf{ClassEval-Completion} \\
\midrule
1.3b-base & 61.96 & 47.47 & 72.86 & 27.20 & 30.30 \\
6.7b-base & 67.76 & 60.32 & 78.29 & 38.50 & 47.47 \\
33b-base  & 67.95 & 64.26 & 79.90 & 47.00 & 56.56 \\
\midrule
1.3b-instruct & 24.10\,(-61.10\%) & 26.82\,(-20.38\%) & 70.90\,(-2.69\%) & 28.30\,(+4.04\%) & 36.62\,(+20.86\%) \\
6.7b-instruct & 46.95\,(-30.71\%) & 23.57\,(-60.93\%) & 77.51\,(-1.00\%) & 41.10\,(+6.75\%) & 57.82\,(+21.80\%) \\
33b-instruct  & 70.57\,(+3.86\%)  & 11.02\,(-82.86\%) & 79.47\,(-0.54\%) & 49.40\,(+5.11\%) & 60.35\,(+6.70\%) \\
\bottomrule
\end{tabular}%
}
\end{table}

\begin{table*}[t]
\centering
\scriptsize
\setlength{\tabcolsep}{3.5pt}
\renewcommand{\arraystretch}{0.95}
\caption{McNemar's test results for base vs.\ instruction-tuned models across benchmarks.
Here, $n_{10}$ denotes cases where the base model succeeds but the instruction-tuned model fails,
while $n_{01}$ denotes cases where the instruction-tuned model succeeds but the base model fails.
For SAFIM, results are reported on the combined API, Block, and Control subsets.
Benchmark abbreviations: HEFIM = HumanEval Infilling, CE-Inf = ClassEval-Infilling,
CE-Comp = ClassEval-Completion, BCB = BigCodeBench-Complete.}
\label{tab:mcnemar_all}
\begin{minipage}[t]{0.49\textwidth}
\centering
\textbf{(a) Qwen2.5-Coder}

\vspace{2pt}
\begin{tabular}{llrrrrc}
\toprule
\textbf{Size} & \textbf{Bench.} & \textbf{$n_{10}$} & \textbf{$n_{01}$} & \textbf{$\chi^2$} & \textbf{$p$-value} & \textbf{Sig.} \\
\midrule
\multirow{5}{*}{1.5B}
& HE-FIM   & 39   & 231  & 135.1  & $<10^{-16}$         & Y \\
& SAFIM    & 1696 & 1805 & 3.3    & $0.068$             & N \\
& CE-Inf   & 262  & 69   & 111.4  & $<10^{-16}$         & Y \\
& CE-Comp  & 22   & 68   & 22.5   & $2.10\times10^{-6}$ & Y \\
& BCB      & 93   & 129  & 5.5    & $0.0188$            & Y \\
\cmidrule(lr){1-7}
\multirow{5}{*}{7B}
& HE-FIM   & 106  & 39   & 30.0   & $4.23\times10^{-8}$ & Y \\
& SAFIM    & 2144 & 1868 & 18.9   & $1.41\times10^{-5}$ & Y \\
& CE-Inf   & 197  & 69   & 60.6   & $6.87\times10^{-15}$& Y \\
& CE-Comp  & 9    & 100  & 74.3   & $<10^{-16}$         & Y \\
& BCB      & 97   & 175  & 21.8   & $3.03\times10^{-6}$ & Y \\
\cmidrule(lr){1-7}
\multirow{5}{*}{14B}
& HE-FIM   & 200  & 10   & 170.1  & $<10^{-16}$         & Y \\
& SAFIM    & 2292 & 952  & 552.7  & $<10^{-16}$         & Y \\
& CE-Inf   & 138  & 45   & 46.3   & $1.04\times10^{-11}$& Y \\
& CE-Comp  & 13   & 128  & 92.2   & $<10^{-16}$         & Y \\
& BCB      & 62   & 123  & 19.5   & $1.03\times10^{-5}$ & Y \\
\cmidrule(lr){1-7}
\multirow{5}{*}{32B}
& HE-FIM   & 44   & 16   & 12.2   & $4.91\times10^{-4}$ & Y \\
& SAFIM    & 2823 & 643  & 1369.9 & $<10^{-16}$         & Y \\
& CE-Inf   & 150  & 46   & 54.1   & $1.88\times10^{-13}$& Y \\
& CE-Comp  & 33   & 56   & 5.4    & $0.0197$            & Y \\
& BCB      & 72   & 130  & 16.1   & $6.06\times10^{-5}$ & Y \\
\bottomrule
\end{tabular}
\end{minipage}
\hfill
\begin{minipage}[t]{0.49\textwidth}
\centering
\textbf{(b) DeepSeek-Coder}

\vspace{2pt}
\begin{tabular}{llrrrrc}
\toprule
\textbf{Size} & \textbf{Bench.} & \textbf{$n_{10}$} & \textbf{$n_{01}$} & \textbf{$\chi^2$} & \textbf{$p$-value} & \textbf{Sig.} \\
\midrule
\multirow{5}{*}{1.3B}
& HE-FIM   & 48   & 103  & 19.3   & $1.11\times10^{-5}$ & Y \\
& SAFIM    & 4453 & 745  & 2643.7 & $<10^{-16}$         & Y \\
& CE-Inf   & 181  & 131  & 7.7    & $0.0055$            & Y \\
& CE-Comp  & 38   & 67   & 7.5    & $0.0063$            & Y \\
& BCB      & 106  & 120  & 0.7    & $0.387$             & N \\
\cmidrule(lr){1-7}
\multirow{5}{*}{6.7B}
& HE-FIM   & 37   & 81   & 15.7   & $7.54\times10^{-5}$ & Y \\
& SAFIM    & 6614 & 474  & 5317.1 & $<10^{-16}$         & Y \\
& CE-Inf   & 101  & 81   & 2.0    & $0.159$             & N \\
& CE-Comp  & 18   & 59   & 20.8   & $5.15\times10^{-6}$ & Y \\
& BCB      & 150  & 127  & 1.7    & $0.186$             & N \\
\cmidrule(lr){1-7}
\multirow{5}{*}{33B}
& HE-FIM   & 54   & 71   & 2.0    & $0.152$             & N \\
& SAFIM    & 9744 & 238  & 9050.8 & $<10^{-16}$         & Y \\
& CE-Inf   & 74   & 63   & 0.7    & $0.393$             & N \\
& CE-Comp  & 13   & 23   & 2.3    & $0.134$             & N \\
& BCB      & 105  & 94   & 0.5    & $0.478$             & N \\
\bottomrule
\end{tabular}
\end{minipage}
\end{table*}

The results indicate several findings: 
\textit{(i) The instruction-tuning tax exists in most cases}. For the FIM tasks, instruct models perform worse than base models except for Qwen2.5-Coder-1.5b and DeepSeek-Coder-33b. The largest performance gap is observed from DeepSeek-Coder-1.3b on HumanEval Infilling, demonstrating a relative degradation of -61.10\%.  However, there are still some exceptions observed in both of the models. Since they come from different model families, the results may be attributed to memorization for smaller models for Qwen2.5-Coder-1.5b and the generalization ability of larger models for DeepSeek-Coder-33b. As the training data for these models are not made open, we are unable to rule out the possibility of memorization for such anomalies observed in HumanEval Infilling (HEFIM). Nevertheless, a prominent evidence from the degradation observed in SAFIM across two model families shows that in the presence of multiple programming languages and task diversity, instruction tuning tax increases steadily with the model size.

\textit{(ii) The magnitude of instruction-tuning tax is not consistent across different model families.} We observe that the performances of the Qwen2.5-Coder and DeepSeek-Coder families do not exhibit similar degrees of relative degradation. For smaller models such as the 1.3/1.5b and 6.7/7b variants, the relative performance degradations in FIM tasks are much more pronounced among the DeepSeek-Coder models. While both of the models are based on decoder-only architecture, they are continually pre-trained and instruction-tuned with different data distributions and ratios, giving rise to different model behaviors, particularly in the degree to which they retain abilities learned through pre-training. Overall, the Qwen2.5-Coder instruct models are instruction-tuned to be more robust in handling infilling and completion tasks compared to DeepSeek-Coder.

\begin{finding}
    {While the instruction-tuning tax exists in most cases, its magnitude varies a lot across different model families. }
\end{finding}

\textit{(iii) Instruct models gain from contextual cues at the expense of their infilling abilities.} Our experiments illustrate this from two aspects. From a task perspective observed in ClassEval, Completion contains to-do descriptions embedded as docstrings, whereas infilling asks for a single missing line with the surrounding codes remained as context. Instruct models performed better than their Base variants in Completion while suffering from the degradations observed in its Infilling counterpart, demonstrating how Instruct models are steered towards a stronger dependence on natural language cues at the expense of its code infilling capability. From a model perspective, we also observe that Instruct models are beneficial to our Completion tasks, all of which are guided with natural language contexts in comments or docstrings.

\begin{finding}
    {Instruct models generally perform better with guidance (in-line docstrings) compared to their Base variants. They tend to improve on understanding natural language cues at the expense of the ability to infill codes with surrounding code context.}
\end{finding}

While aggregate metrics such as Pass@1 quantify performance differences, they do not reveal whether these differences arise from consistent behavioral shifts or from a small number of cases. To further examine whether instruction tuning induces systematic changes in model behavior, we apply McNemar’s test to statistically compare the performances of base and instruction-tuned models on a per-instance basis: a significance in the test implies that the performances between two models (i.e., base and instruct) are distinctive. The results are reported in Table \ref{tab:mcnemar_all}.

\textit{(iv) The effect of model size on instruction-tuning tax is model-specific.} The McNemar’s test results reveal a clear contrast between model families. For Qwen2.5-Coder, differences between base and instruction-tuned models are consistently significant across most benchmarks and model sizes, indicating a stable behavioral shift induced by instruction tuning. In contrast, DeepSeek-Coder shows weaker and diminishing significance as model size increases, suggesting greater overlap between base and instruct outputs in larger models. Notably, SAFIM remains highly significant across all models, with a consistent imbalance where base models outperform instruction-tuned variants, indicating a persistent degradation on structured infilling tasks.

\begin{finding}
{There is no guarantee of eliminating instruction-tuning tax with increasing model sizes as its effect varies with different model families.
}
\end{finding}

\subsection{RQ2: What are the underlying reasons for the performance differences implied by the change in models' behavior?}

To better understand the behavioral changes underlying the performance differences observed in RQ1, we analyze how base and instruction-tuned models differ from three perspectives. First, we compare the sets of problems solved by each model variant to identify where their correct predictions overlap and diverge. Second, we inspect sampled failed cases to understand the concrete failure modes introduced by instruction tuning. Third, we validate whether these observed failure patterns reflect systematic behavioral differences across the full set of raw model outputs.

\begin{figure*}[tbp]
    \centering
    \includegraphics[width=1\textwidth]{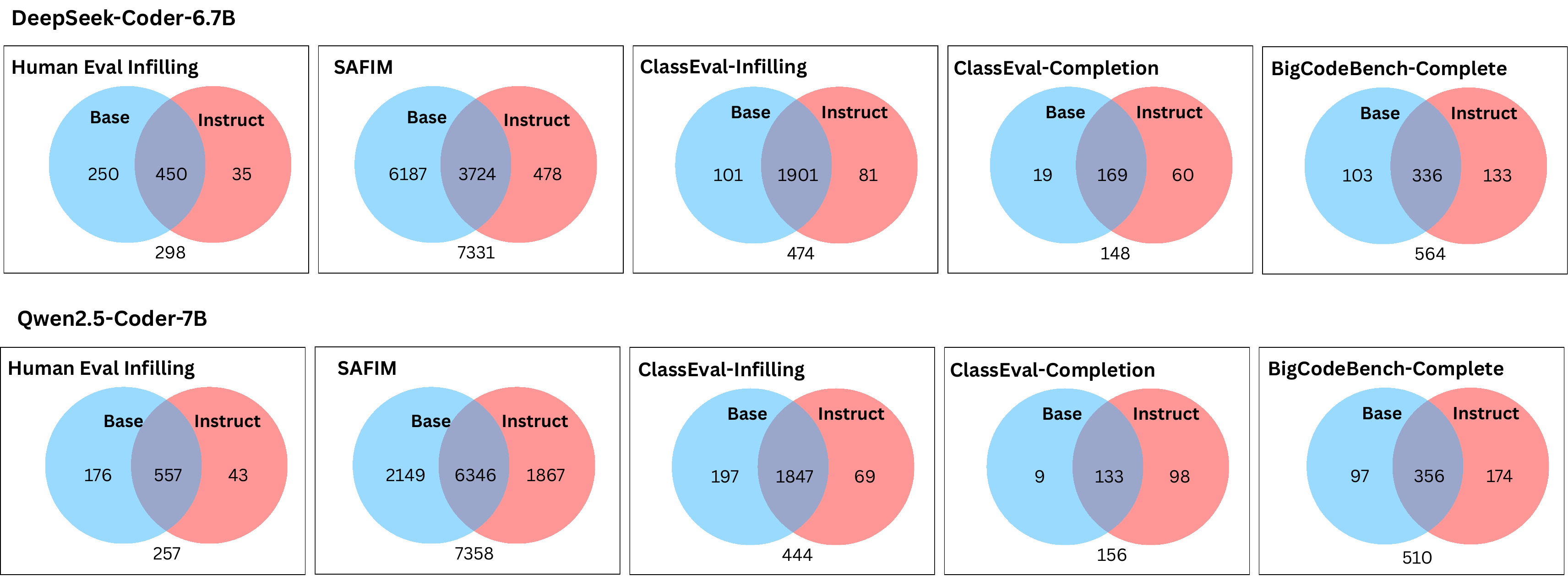}
    \vspace{-10pt}  
    \caption{The sets of problems solved by DeepSeek-Coder and Qwen2.5-Coder illustrated in Venn diagrams.}
    \label{fig:venndiagram}
    \vspace{-10pt}  
\end{figure*}

We first compare the distributions of solved problems using Venn diagrams, as shown in Figure~\ref{fig:venndiagram}. The number of problems solved by both base and instruction-tuned models is larger than the number solved exclusively by either model, indicating substantial commonality in their capabilities for both DeepSeek-Coder-6.7B and Qwen2.5-Coder-7B. However, the exclusively solved problems reveal task-specific differences. For FIM tasks, both base models solve more problems exclusively than their instruction-tuned counterparts. In contrast, for completion tasks evaluated through BigCodeBench and ClassEval, instruction-tuned models show stronger exclusive advantages.

\begin{finding}
{While base and instruction-tuned models share a substantial overlap in their correct answers, their exclusive successes differ by task. Base models exclusively solve more FIM problems, whereas instruction-tuned models exclusively solve more completion problems.}
\end{finding}

\paragraph{\textbf{RQ2.1: What failure patterns emerge after instruction tuning?}}

To inspect how models fail after instruction tuning, we conduct a sampled failure-mode analysis. We randomly sample 50 failed instances from each base and instruction-tuned model across Code Infilling benchmarks, including HumanEval Infilling, ClassEval Infilling, and SAFIM, and Completion benchmarks, including ClassEval Completion and BigCodeBench Complete. Two authors manually categorize the sampled failures, achieving an agreement score of 0.8234 measured by Cohen's kappa ($\kappa$)~\cite{mchugh2012interrater}, which indicates almost perfect agreement.

\begin{figure*}[tbp]
    \centering
    \includegraphics[width=1\textwidth]{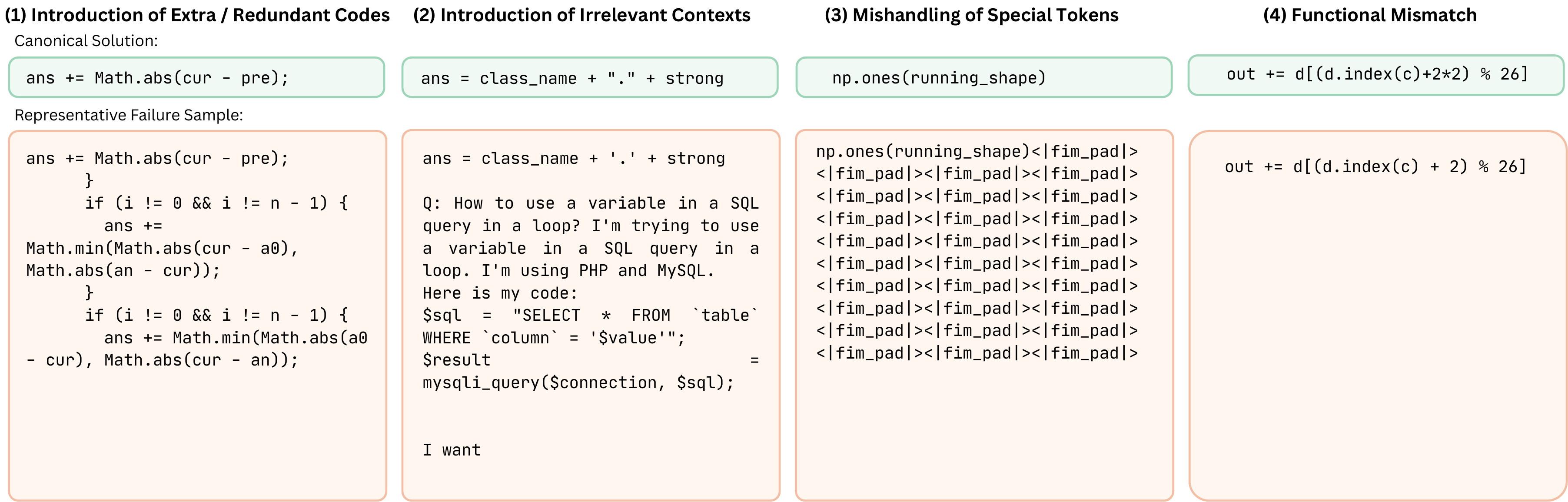}
    \caption{Representative samples of the failure categories.}
    \label{fig:failure_rep}
\end{figure*}

We categorize the sampled failed cases into four main categories. Representative examples of each category are shown in Figure~\ref{fig:failure_rep}.

\begin{enumerate}
    \item Introduction of Extra/Redundant Codes: The model generates unnecessary codes, such as unintended continuations, additional function definitions, or irrelevant snippets.
    \item Introduction of Irrelevant Contexts: The response contains excessive natural language intermixed with code, making resulting code unparsable and thus requiring non-trivial postprocessing.
    \item Mishandling of Special Tokens: The model outputs special tokens (e.g., \verb+<|fim_pad|>+) that should not appear in responses.
    \item Functional Mismatch: Although syntactically correct, the response fails evaluation due to logical or semantic mismatches, resulting in a failed evaluation result. The cases of incomplete generations are also included under this category.
\end{enumerate}

The results of the manual categorization are shown in Table~\ref{tab:ret-failurecategory-infilling} for infilling tasks and Table~\ref{tab:ret-failurecategory-completion} for completion tasks.

\begin{table*}[t]
\centering
\scriptsize
\caption{Distribution of failure categories obtained from manual analyses on infilling benchmarks.}
\label{tab:ret-failurecategory-infilling}
\resizebox{\textwidth}{!}{
\begin{tabular}{l|cc|cc|cc|cc|cc|cc}
\toprule
\textbf{} 
& \multicolumn{4}{c|}{\textbf{HumanEval Infilling}} 
& \multicolumn{4}{c|}{\textbf{ClassEval Infilling}} 
& \multicolumn{4}{c}{\textbf{SAFIM}} \\
\cmidrule(lr){2-5} \cmidrule(lr){6-9} \cmidrule(lr){10-13}
& \multicolumn{2}{c|}{\textbf{Qwen2.5-Coder-7B}} 
& \multicolumn{2}{c|}{\textbf{DSC-6.7B}}
& \multicolumn{2}{c|}{\textbf{Qwen2.5-Coder-7B}} 
& \multicolumn{2}{c|}{\textbf{DSC-6.7B}}
& \multicolumn{2}{c|}{\textbf{Qwen2.5-Coder-7B}} 
& \multicolumn{2}{c}{\textbf{DSC-6.7B}} \\
\cmidrule(lr){2-3} \cmidrule(lr){4-5}
\cmidrule(lr){6-7} \cmidrule(lr){8-9}
\cmidrule(lr){10-11} \cmidrule(lr){12-13}
\textbf{Category} 
& \textbf{base} & \textbf{instruct} 
& \textbf{base} & \textbf{instruct}
& \textbf{base} & \textbf{instruct} 
& \textbf{base} & \textbf{instruct}
& \textbf{base} & \textbf{instruct} 
& \textbf{base} & \textbf{instruct} \\
\midrule
Introduction of Extra/Redundant Codes 
& 10\% & 48\% & 42\% & 94\%
& 2\%  & 4\%  & 0\%  & 4\%
& 22\% & 22\% & 10\% & 58\% \\

Introduction of Irrelevant Contexts    
& 0\%  & 0\%  & 0\%  & 4\%
& 0\%  & 4\%  & 0\%  & 2\%
& 0\%  & 6\%  & 0\%  & 0\%  \\

Mishandling of Special Tokens          
& 0\%  & 0\%  & 0\%  & 0\%
& 0\%  & 0\%  & 0\% & 0\%
& 0\%  & 2\%  & 0\%  & 0\%  \\

Functional Mismatch                    
& 80\% & 52\% & 58\% & 2\%
& 98\% & 92\% & 100\% & 94\%
& 78\% & 70\% & 90\% & 42\% \\
\bottomrule
\end{tabular}}
\end{table*}

\begin{table*}[t]
\centering
\scriptsize
\caption{Distribution of failure categories obtained from manual analyses on completion benchmarks.}
\label{tab:ret-failurecategory-completion}
\resizebox{\textwidth}{!}{
\begin{tabular}{l|cc|cc|cc|cc}
\toprule
\textbf{} 
& \multicolumn{4}{c|}{\textbf{ClassEval Completion}} 
& \multicolumn{4}{c}{\textbf{BigCodeBench Complete}} \\
\cmidrule(lr){2-5} \cmidrule(lr){6-9}
& \multicolumn{2}{c|}{\textbf{Qwen2.5-Coder-7B}} 
& \multicolumn{2}{c|}{\textbf{DSC-6.7B}}
& \multicolumn{2}{c|}{\textbf{Qwen2.5-Coder-7B}} 
& \multicolumn{2}{c}{\textbf{DSC-6.7B}} \\
\cmidrule(lr){2-3} \cmidrule(lr){4-5}
\cmidrule(lr){6-7} \cmidrule(lr){8-9}
\textbf{Category} 
& \textbf{base} & \textbf{instruct} 
& \textbf{base} & \textbf{instruct}
& \textbf{base} & \textbf{instruct} 
& \textbf{base} & \textbf{instruct} \\
\midrule
Introduction of Extra/Redundant Codes 
& 6\%  & 48\% & 34\% & 98\%
& 0\%  & 6\%  & 16\% & 22\% \\

Introduction of Irrelevant Contexts    
& 52\% & 6\%  & 46\% & 2\%
& 4\%  & 0\%  & 0\%  & 0\%  \\

Mishandling of Special Tokens          
& 0\%  & 0\%  & 0\%  & 0\%
& 0\%  & 0\%  & 0\%  & 0\%  \\

Functional Mismatch                    
& 42\% & 44\% & 20\% & 0\%
& 96\% & 94\% & 84\% & 78\% \\
\bottomrule
\end{tabular}}
\end{table*}

\begin{table*}[t]
\centering
\scriptsize
\setlength{\tabcolsep}{2pt}
\renewcommand{\arraystretch}{0.95}
\caption{Behavioral metrics with failure-conditioned values shown in parentheses. Each cell reports overall weighted performance first, followed by the corresponding failure-conditioned value in parentheses. Infilling uses dataset-size weights (HEFIM 1,033, SAFIM-Weighted 17,720, ClassEval-Infilling 2,557); completion uses ClassEval-Completion 396 plus BigCodeBench-Complete 1,140.}
\label{tab:failure-rates-task-weighted-overall-with-failure-only}
\begin{minipage}[t]{0.48\textwidth}
\centering
\textbf{Qwen2.5-Coder}

\resizebox{\linewidth}{!}{%
\begin{tabular}{lcccccc}
\toprule
Model & \multicolumn{3}{c}{Infilling} & \multicolumn{3}{c}{Completion} \\
 & CTR$\uparrow$ & NL$\downarrow$ & Def/Cls$\downarrow$ & CTR$\uparrow$ & NL$\downarrow$ & Def/Cls$\downarrow$ \\
\midrule
1.5B & 90.18 (38.44) & 0.66 (2.22) & 2.10 (10.09) & 84.46 (52.29) & 8.46 (20.55) & 2.80 (8.16) \\
7B & 92.93 (12.71) & 0.16 (2.34) & 0.56 (7.60) & 89.63 (70.38) & 8.53 (24.02) & 2.93 (10.97) \\
14B & 96.96 (11.15) & 0.26 (7.74) & 0.10 (3.23) & 93.97 (90.88) & 5.40 (16.01) & 7.49 (18.65) \\
32B & 97.08 (9.90) & 0.14 (7.04) & 0.12 (4.01) & 89.75 (80.79) & 2.93 (14.39) & 4.36 (16.97) \\
\cmidrule(lr){1-7}
1.5B-Instruct & 89.74 (7.33) & 0.45 (3.20) & 1.65 (7.26) & 86.51 (78.87) & 8.20 (12.61) & 9.18 (14.06) \\
7B-Instruct & 88.72 (16.49) & 2.50 (15.39) & 2.70 (10.37) & 89.03 (78.60) & 9.31 (13.18) & 8.27 (15.00) \\
14B-Instruct & 91.78 (20.50) & 2.69 (24.97) & 1.47 (9.01) & 90.73 (75.07) & 6.12 (7.63) & 2.73 (3.53) \\
32B-Instruct & 89.48 (16.65) & 2.27 (17.52) & 1.14 (9.29) & 86.21 (77.19) & 14.00 (17.33) & 9.83 (15.52) \\
\bottomrule
\end{tabular}%
}
\end{minipage}
\hfill
\begin{minipage}[t]{0.48\textwidth}
\centering
\textbf{DeepSeek-Coder}

\resizebox{\linewidth}{!}{%
\begin{tabular}{lcccccc}
\toprule
Model & \multicolumn{3}{c}{Infilling} & \multicolumn{3}{c}{Completion} \\
 & CTR$\uparrow$ & NL$\downarrow$ & Def/Cls$\downarrow$ & CTR$\uparrow$ & NL$\downarrow$ & Def/Cls$\downarrow$ \\
\midrule
1.3B & 90.63 (3.77) & 0.16 (2.11) & 1.41 (8.12) & 79.50 (71.80) & 5.14 (14.68) & 18.62 (21.42) \\
6.7B & 93.72 (4.11) & 0.16 (2.68) & 0.79 (6.48) & 82.06 (80.35) & 5.66 (8.61) & 15.04 (18.88) \\
33B & 94.92 (6.77) & 0.21 (5.14) & 1.03 (6.36) & 80.16 (77.52) & 3.32 (11.86) & 17.38 (25.40) \\
\cmidrule(lr){1-7}
1.3B-Instruct & 62.60 (14.39) & 6.64 (17.60) & 13.33 (16.03) & 85.80 (85.49) & 11.52 (14.11) & 9.18 (9.99) \\
6.7B-Instruct & 58.38 (9.09) & 2.95 (4.51) & 14.63 (14.07) & 79.68 (78.43) & 4.36 (7.49) & 24.02 (26.53) \\
33B-Instruct & 54.64 (10.37) & 5.17 (6.41) & 17.41 (15.31) & 79.20 (78.84) & 3.65 (5.56) & 23.18 (24.62) \\
\bottomrule
\end{tabular}%
}
\end{minipage}
\vspace{2pt}

\end{table*}

In the HumanEval Infilling benchmark, extra or redundant code accounted for 10\% of failures in the base Qwen2.5-Coder model and 48\% in the instruct model, while for DeepSeek-Coder, it reached 42\% in the base model and 94\% in the instruct model, with an additional 4\% of failures due to the introduction of irrelevant contexts. A similar pattern is observed in ClassEval Infilling. For Qwen2.5-Coder-7B, extra or redundant code increases from 2\% in the base model to 4\% in the instruct model, with the majority of failures still dominated by functional mismatches (98\% → 92\%). For DeepSeek-Coder-6.7B, we observe 10\% of failure cases of Base related to mishandling of special tokens while redundant and irrelevant contexts arises in the failure cases of Instruct. Across benchmarks, although most failures arise from functional mismatches, instruction-tuned models consistently exhibit a higher proportion of extra or redundant codes, suggesting that instruct models tend to fail these jobs by behaving verbose. On SAFIM using Qwen2.5-Coder-7B, the base model introduced extra or redundant code in 22\% of failures, with all remaining failures classified as functional mismatches, whereas the instruct model has additional errors with irrelevant contexts and mishandling of special tokens. On the other hand, on DeepSeek-Coder-6.7B, the base model exhibited 10\% of failures due to extra or redundant code, while that of the instruct model increased to 58\%. 

We observe similar behavioral shifts in the case of completion tasks. In ClassEval Completion, the primary failure mode of Qwen2.5-Coder-7B shifts from irrelevant contexts (52\%) for the base model to redundant codes (48\%) for the instruct model. DeepSeek-Coder-6.7B shows an even larger change: its base model distributes failures across irrelevant contexts (46\%), while the instruct model significantly fails through extra or redundant code (98\%). On the other hand, BigCodeBench Complete remains dominated by functional mismatches for both model families, with failure modes shifting towards redundant codes after instruction tuning. Overall, although functional mismatch is still the most common failure category, instruction-tuned models often show a higher proportion of extra or redundant code, suggesting that instruction tuning can make failures more verbose rather than purely functional.

\begin{finding}
{Manual analyses of sampled failure cases show that most failures remain functional mismatches, but instruction tuning shifts a larger share of errors toward verbosity-related failure modes, especially extra or redundant code and irrelevant natural language.}
\end{finding}

\paragraph{\textbf{RQ2.2: Can we quantify these failure modes at scale?}} The manual analysis suggests that instruction-tuned models often fail by mixing otherwise plausible code with explanatory or structurally extraneous text. To test whether this pattern is systematic rather than anecdotal, we compute three rule-based behavioral metrics over the full set of raw generations: Code-to-Token Ratio (CTR), natural-language rate (NL), and extra function/class declaration rate (Def/Cls). 

For CTR, each output is processed using a parser-free lexical tokenizer based on Python's standard tokenizer. We use this tokenizer only as a lightweight lexical approximation, rather than as a Python-specific parser, because some benchmarks contain partial, malformed, mixed-format, or non-Python outputs. The denominator includes all non-comment lexical tokens, excluding pure formatting or boundary tokens emitted by the tokenizer, such as line breaks, indentation changes, encoding markers, and end-of-sequence markers. The numerator counts code-like tokens, including operator tokens, numeric literals, string literals, and word-like identifier/keyword tokens. Because this category also captures many alphabetic words that may appear in natural-language text, CTR should be interpreted as a lexical code-density measure rather than a perfect separator between code and prose. We report the mean CTR over all samples in each dataset.

NL is reported as the percentage of samples flagged as containing natural-language text. A sample is marked positive if it matches a predefined set of discourse markers, including ``here'', ``this'', ``explanation'', ``because'', ``therefore'', ``note'', ``please'', ``you'', ``we'', ``the following'', ``in summary'', ``overall'', ``i think'', and ``let me'', or if the output contains at least eight alphabetic words and at most two code-punctuation characters. Def/Cls is reported as the percentage of samples containing at least one line that matches a \texttt{def}- or \texttt{class}-like declaration pattern. These metrics are directly motivated by the preceding manual analysis, except that special-token errors are omitted because they are concentrated in the Qwen family and do not represent a stable cross-family trend. Table~\ref{tab:failure-rates-task-weighted-overall-with-failure-only} reports the overall task-weighted value first, with the corresponding failure-conditioned value in parentheses, where the latter is computed only over failed outputs. This allows us to compare both aggregate generation behavior and the composition of failed outputs.

At the aggregate level, the distinctions between the models' behavior metrics in infilling and completion tasks are apparent. Across both model families, infilling is substantially more stable than completion, as reflected by its generally higher CTR and lower NL and Def/Cls rates. This suggests that suffix-constrained generation provides a stronger structural prior, keeping outputs concise and code-focused. Completion, in contrast, is more weakly constrained: because the model must infer both what to generate and where to stop, it shows a greater tendency to produce explanatory text or structurally unnecessary code.

The failure-conditioned values further reveal that this gap is not only quantitative but also qualitative. Once an infilling output fails, its code-to-token ratio often collapses sharply. For example, Qwen-14B drops from 96.96 to 11.15 and DeepSeek-Coder-6.7B from 93.72 to 4.11, suggesting that many infilling failures are dominated by under-generated or non-code-like outputs. Completion failures behave differently. For several stronger models, failed completion outputs remain highly code-dense, such as Qwen-14B (93.97 to 90.88) and DeepSeek-Coder-6.7B (82.06 to 80.35), while natural-language markers and extra definitions/classes increase markedly. This indicates that completion failures often still contain substantial code, but are excessively generated rather than merely too short.

This differential view also helps pinpoint how instruction tuning changes behavior within each model family. For Qwen, instruction tuning is accompanied by especially large increases in infilling NL among failed outputs, most notably for 14B-Instruct (2.69 to 24.97), indicating that verbosity penetrates even suffix-constrained settings once the model deviates from the target span. For DeepSeek-Coder, the behavioral shift is particularly visible in completion, where failed outputs remain consistently high in Def/Cls, suggesting a stronger tendency to over-generate with unnecessary code. Taken together, these results show that instruction tuning does not merely change how often models fail; it also changes \emph{how} they fail. Infilling failures are more often low-code, whereas completion failures are more often code-heavy but excessively generated.

Viewed more broadly, these results suggest that the primary robustness gap arises from the structural difference between unconstrained continuation and suffix-constrained infilling, rather than purely from model-specific factors. At the same time, instruction tuning systematically alters generation behavior across both settings, introducing a divergence between functional correctness and generation fidelity. Even when outputs generally remain executable or code-dense, the failures reveal non-trivial redundant contexts that persist despite best-effort post-processing.

\begin{finding}
{The behavioral metrics further show that the verbosity trend observed in sampled failures is systematic across outputs. Instruction-tuned models are more likely to introduce natural language or unnecessary definitions/classes, especially in weakly constrained completion tasks, while suffix-constrained infilling tasks generally keeps outputs more code-focused.}
\end{finding}

\subsection{RQ3: How does model performance evolve during the fine-tuning process from Base to Instruct?}

To examine how performance evolves during instruction tuning, we evaluate intermediate checkpoints from the Magicoder pipeline applied to Qwen2.5-Coder-7B-Base. Different from RQ1 and RQ2, we additionally include instruction-following tasks in this experiment to observe whether and when instruction-following ability emerges during the tuning process. As these tasks are evaluated with the instruct-model chat template, the untuned base model is not evaluated in this setting, and thus its entries are reported as NA in Table~\ref{tab:rq3_checkpoints}. We focus on Qwen2.5-Coder because, from RQ1, its base and instruct variants exhibit a more consistent and statistically significant behavioral difference across benchmarks than DeepSeek-Coder, making it a more suitable target for analyzing how instruction tuning reshapes capability over time. We further select the 7B model as a representative mid-scale setting that is large enough to exhibit the main trade-offs observed in RQ1, while still remaining computationally manageable for repeated checkpoint-based fine-tuning and evaluation. We note that the tuning process is conducted at a relatively small scale and is not intended to recover the full proprietary tuning trajectories of official instruct models such as Qwen2.5-Coder or DeepSeek-Coder. Instead, our setup should be interpreted as a controlled sensitivity test: by applying incremental amounts of publicly available instruction-tuning data, we examine how a base code model responds across different task types when exposed to instruction tuning.

Across intermediate checkpoints, the effect of instruction tuning varies with tasks and is non-monotonic. This volatility is already visible within the OSS-Instruct-75k stage: with only 5\% of the data, the model attains 81.7 on HumanEval(+) and 78.4 on MBPP(+), which is already as strong as, or stronger than, many later checkpoints. Subsequent tuning does not consistently preserve these gains, suggesting that a small amount of instruction data is sufficient to elicit instruction-following behavior, while additional tuning mainly reshapes how that behavior interacts with other tasks. The second 110k stage further amplifies this instability, as several benchmarks shift direction again instead of continuing the trends established in the 75k phase. Rather than producing a uniform progression from weaker to stronger checkpoints, tuning leads to repeated rises and drops across benchmarks, suggesting that capabilities across different task types are being reallocated rather than uniformly improved when fine-grained intermediate checkpoints are evaluated.

A more stable pattern emerges when the results are grouped by task type. For infilling tasks, the long-run effect is generally negative, although the decline is not strictly smooth. SAFIM decreases from 47.95 in the base model to values mostly in the 44--47 range during the 75k stage, and reaches a low of 37.44 at the 110k-40\% checkpoint before partially recovering. HumanEval Infilling (HEFIM) is more resilient, remaining close to the base model in early checkpoints and even slightly exceeding it at 5\% and 10\% of the 75k phase, but later checkpoints again trend downward and settle around 70.09 at 110k-100\%. ClassEval-Infilling behaves similarly: its scores fluctuate within a narrow band but show no sustained improvement over the base model, ending below the initial 79.40. Taken together, these results suggest that instruction tuning tends to weaken local code insertion and infilling-style generation, even though short-lived recoveries appear at intermediate checkpoints.

\begin{table*}[t]
    \centering
    \caption{Evaluation results of intermediate checkpoints obtained from instruction-tuning Qwen2.5-Coder-7B via the Magicoder pipeline (OSS-Instruct-75k followed by Evol-Instruct-110k). Brackets in HE(+) and MBPP(+) indicate the evaluation results with extra test cases.}
    \label{tab:rq3_checkpoints}
    \resizebox{\textwidth}{!}{%
    \begin{tabular}{@{}l|ccc|cc|cccc@{}}
    \toprule
    \textbf{} 
    & \multicolumn{3}{c|}{\textbf{Fill in the Middle}} 
    & \multicolumn{2}{c|}{\textbf{Completion}} 
    & \multicolumn{4}{c}{\textbf{Instruction Following}} \\
    \cmidrule(lr){2-4} \cmidrule(lr){5-6} \cmidrule(l){7-10}
    \textbf{} 
    & \textbf{HEFIM} 
    & \textbf{SAFIM} 
    & \textbf{CE-Inf} 
    & \textbf{BCB-Comp} 
    & \textbf{CE-Comp} 
    & \textbf{HE(+)} 
    & \textbf{MBPP(+)} 
    & \textbf{DS1000} 
    & \textbf{BCB-Inst} \\
    \midrule
    \textbf{Qwen2.5-Coder-7B} 
    & 70.95 & 47.95 & 79.94 
    & 39.74 & 35.85 
    & NA & NA & NA & NA \\
    \midrule
    \textbf{75k Instruct} & \multicolumn{9}{c}{} \\
    \midrule
    5\%   
    & 72.22 & 46.50 & 78.37 
    & 38.00 & 50.51 
    & 81.7(75.6) & 78.4(63.4) & 12.70 & 36.10 \\
    10\%  
    & 71.25 & 48.88 & 78.92 
    & 37.40 & 54.04 
    & 76.8(70.7) & 77.2(63.7) & 12.90 & 35.70 \\
    20\%  
    & 67.96 & 45.71 & 78.26 
    & 37.90 & 57.58 
    & 79.3(73.8) & 76.2(62.2) & 15.50 & 33.20 \\
    40\%  
    & 69.99 & 46.67 & 78.69 
    & 39.30 & 59.59 
    & 78.7(70.7) & 76.2(63.2) & 9.30  & 35.80 \\
    60\%  
    & 70.57 & 40.65 & 78.06 
    & 33.20 & 49.49 
    & 78.7(74.4) & 76.9(65.7) & 13.80 & 35.10 \\
    80\%  
    & 68.73 & 44.93 & 78.02 
    & 34.90 & 53.54 
    & 81.1(77.4) & 77.7(62.9) & 13.30 & 36.20 \\
    100\% 
    & 71.35 & 46.90 & 78.22 
    & 38.20 & 59.34 
    & 78.0(74.4) & 77.2(64.2) & 14.60 & 35.10 \\
    \midrule
    \textbf{110k Instruct} & \multicolumn{9}{c}{} \\
    \midrule
    20\%  
    & 69.51 & 50.45 & 77.55 
    & 31.90 & 54.80 
    & 81.7(76.8) & 76.9(65.4) & 33.20 & 34.30 \\
    40\%  
    & 70.47 & 37.44 & 77.71 
    & 31.40 & 57.07 
    & 79.3(73.2) & 76.7(63.4) & 35.60 & 35.30 \\
    60\%  
    & 68.34 & 48.17 & 77.43 
    & 32.90 & 57.07 
    & 82.3(76.2) & 76.9(65.2) & 37.40 & 35.30 \\
    80\%  
    & 71.25 & 50.93 & 77.51 
    & 32.30 & 58.83 
    & 82.3(77.4) & 76.7(64.7) & 36.90 & 33.30 \\
    100\% 
    & 70.09 & 45.25 & 77.12 
    & 27.90 & 59.34 
    & 81.7(78.0) & 76.4(63.4) & 37.90 & 33.30 \\
    \bottomrule
    \end{tabular}
    }
\end{table*}

The strongest tuning effects appear on ClassEval-Completion, BigCodeBench-Complete, and DS1000-Instruct, but importantly not in the same direction. DS1000-Instruct shows the clearest late-stage gain: it starts at only 12.70--15.50 throughout the 75k checkpoints, then jumps sharply to 33.20 at the start of the 110k phase and reaches 37.90 by 110k-100\%. ClassEval-Completion also exhibits large sensitivity to tuning, but here the effect is broadly positive: from a base score of 35.85, nearly all checkpoints improve substantially, peaking at 59.59 in the 75k-40\% checkpoint and remaining above the base model even after subsequent fluctuations. In contrast, BigCodeBench-Complete deteriorates the most in the tuning process: every intermediate checkpoint underperforms the base model, and the score deteriorates further in the 110k phase, falling as low as 27.90. Nevertheless, the divergence observed among the two completion tasks suggests that code completion may not be a single homogeneous capability under the Magicoder tuning experiment: class-scoped completion benefits from tuning, whereas more open-ended and singular completion tasks appear much more vulnerable to the tuning effect.

To observe a long-run averaged trend, we further compute grouped relative changes with respect to the corresponding base averages. For each task category (infilling, instruction following, or completion), we first average the benchmark scores within a stage, and then compute the relative change of that stage average. Specifically, we derive the average performance of a single task category/type by $\frac{\Sigma_{b\in B_{task}} \Sigma_{s\in S} P_{b,s}}{|B_{task}||s|}$ where $P_{b,s}$ is the Pass@1 of a specific benchmark ($b$) at a single stage of fine-tuning ($s$). The relative delta is then computed against the corresponding base average. For instruction-oriented tasks, where no base counterpart exists, we compare the 110k stage against the 75k stage. The statistics in Table \ref{tab:rq3-grouped-relative} show that infilling-oriented tasks exhibit consistent degradation, declining by 1.97\% in the 75k phase and 2.25\% in the 110k phase relative to the base model. Completion-oriented tasks remain above the base model on average, but their relative gain diminishes from 6.86\% in the 75k stage to 2.65\% in the 110k stage, indicating weakening benefits as tuning progresses. For instruction-oriented tasks, although direct base-relative comparison is not available, the family-level average increases by 11.68\% from the 75k stage to the 110k stage, suggesting continued strengthening of instruction-conditioned behavior. 

\begin{table}[t]
\centering
\footnotesize
\caption{Stage-level grouped averages with relative changes. Infilling and Completion deltas are computed relative to the corresponding base family averages. Instruction $\Delta^{\dagger}$ is computed relative to the 75k instruction-family average, since the base model was not evaluated on the instruction-oriented benchmarks.}
\label{tab:rq3-grouped-relative}
\begin{tabular}{lcccccc}
\toprule
\textbf{Stage} & \textbf{Infilling Avg.} & \textbf{Infilling $\Delta$} & \textbf{Instruction Avg.} & \textbf{Instruction $\Delta^{\dagger}$} & \textbf{Completion Avg.} & \textbf{Completion $\Delta$} \\
\midrule
Base  & 66.10 & 0.00\%  & NA    & NA      & 42.98 & 0.00\% \\
75k   & 64.80 & -1.97\% & 51.19 & 0.00\%  & 45.93 & +6.86\% \\
110k  & 64.61 & -2.25\% & 57.17 & +11.68\% & 44.11 & +2.65\% \\
\bottomrule
\end{tabular}
\end{table}

\begin{finding}
{Instruction tuning affects capabilities unevenly across benchmarks: it tends to help explicit instruction-oriented tasks, but often destabilizes or degrades infilling and realistic completion tasks. Although checkpoint-level results are volatile, the aggregate pattern is consistent: later tuning stages reallocate capability rather than uniformly improving all forms of code generation.}
\end{finding}



\section{Discussion}

\subsection{Implication for findings}

\textbf{For researchers.}
These findings offer insights into how instruction tuning interacts with model design and data, guiding future research on improving code LLMs.

\textit{(1) Investigate model-specific instruction-tuning tax.}
Based on the results of RQ1 (Findings 1--3), the instruction-tuning tax is not equally expressed across model families. While aggregate results show degradation in many settings, the McNemar tests further suggest that instruction tuning induces a much more consistent behavioral shift in Qwen than in DeepSeek-Coder, where the tax becomes weaker or even statistically indistinguishable at larger scales. This indicates that the effect of instruction tuning is shaped not only by task type, but also by model-specific factors such as architecture, pretraining distribution, and post-training sensitivity. Future work should therefore move beyond treating instruction-tuning tax as a universal phenomenon, and instead examine the factors that contribute to instruction-tuning tax in order to minimize the trade-off.

\textit{(2) Devise more robust training strategies to reduce instruction-tuning tax.}
Our findings in RQ2 suggest that the instruction-tuning tax is not only a performance phenomenon, but also a behavioral one. Across both the manual failure analysis and the dataset-level generation-discipline metrics, instruction-tuned models in Flow mode show a clear tendency toward verbosity, including extra or redundant code, irrelevant natural-language continuation, lower code-to-token ratio, and higher non-code generation rates. Since these behaviors are closely tied to the assistant-style generation patterns encouraged by instruction tuning itself, future work should explore training strategies that explicitly limit the penetration of verbosity into structurally constrained settings such as infilling. This points to a promising direction for SWE and CodeLLM researchers: rather than treating instruction-tuning tax as an unavoidable trade-off, future methods could mitigate it from a behavioral perspective by preserving concise, boundary-aware continuation during post-training.

\textbf{For practitioners.}
From a practitioner’s perspective, the results carry important implications for model selection and deployment.  

\textit{(1) Deployment safeguards should be guided by model-specific failure modes.}
Our analyses show that different model variants do not merely differ in overall accuracy; they also fail in systematically different ways. In particular, base and instruction-tuned models, as well as different model families such as Qwen2.5-Coder and DeepSeek-Coder, exhibit different compositions of failure modes, in terms of our taxonomy and behavior-based evaluation metrics in RQ2. This suggests that practitioners should not choose models solely based on aggregate benchmark scores, but also consider the kinds of output errors they are most likely to encounter in deployment. Such distinctions are especially important when designing post-processing or validation pipelines to assure generation fidelity and usefulness, since different failure modes may require different mitigation strategies. For example, rule-based cleanup may be effective for trimming redundant continuations or removing formatting artifacts by fixed identifiers, while other cases may require stronger semantic verification. As neither rule-based post-processing nor LLM-based verification can perfectly eliminate all output mismatches, understanding the dominant failure patterns of each model can help practitioners prioritize the most effective and efficient safeguards for their target use cases.

\textit{(2) The design of deployment system architectures remains an open problem.}
Our findings suggest that deployment should not be reduced to a simple choice between base and instruction-tuned models. While a straightforward strategy is to assign (route) flow-oriented tasks such as infilling and completion to base models, and command-oriented tasks such as instruction following to instruct models, the behavioral differences observed in our study such as the advantages in completion tasks as a result of instruction tuning indicate that this view may still be incomplete. An open problem for future work is how deployment architectures should be designed to account for these differences in generation behavior and failure patterns across different task modes especially when human factors such as a developer's mental model are taken into consideration.

\subsection{Threats to validity}
\textbf{Internal Threats.}
\textit{(1) Data leakage.}
One potential threat to validity is data leakage, since our evaluation relies on public benchmarks and the full pre-training and post-training corpora of official CodeLLMs are not fully disclosed. Therefore, we cannot completely rule out the possibility that some benchmark instances were seen during model training, or that Base and Instruct variants had unequal exposure to benchmark-related data. To mitigate this threat, we evaluate models on a diverse set of benchmarks drawn from different sources, including human-authored problems, StackOverflow-derived tasks, GitHub- and Codeforces-based infilling tasks, and benchmark variants with different construction strategies. In addition, RQ3 provides a more controlled setting by applying the same public instruction-tuning pipeline and datasets to a single base model, which allows us to observe capability shifts under known tuning data. Nevertheless, benchmark contamination remains a limitation of studies using public CodeLLMs and public evaluation datasets.

\textit{(2) Model selection.}
Our study focuses on the Qwen2.5-Coder and DeepSeek-Coder families because they best satisfy the requirements of our experimental design. Both families provide matched Base and Instruct variants across multiple parameter scales, which is necessary for analyzing instruction-tuning tax with respect to model family and model size. Their open-source availability also enables reproducible evaluation and, in the case of our RQ3 design, controlled fine-tuning experiments. However, these two model families may not represent all CodeLLMs, especially models trained with different pre-training objectives, instruction-tuning recipes, alignment methods, or proprietary data. Therefore, our goal is not to claim universal generalizability to all CodeLLMs, but to provide controlled empirical evidence that instruction tuning can introduce task-dependent trade-offs in representative open-source CodeLLM families.

\textit{(3) Instruction-tuning setup.}
Another internal threat concerns the scale and scope of our replicated instruction-tuning process in RQ3. Although we follow the public Magicoder pipeline and evaluate multiple intermediate checkpoints, our tuning setup is still smaller and more controlled than the full post-training processes used to produce official instruct models. Therefore, the checkpoint-level findings in RQ3 should not be interpreted as fully generalizable to all possible instruction-tuning processes or proprietary post-training pipelines. Instead, RQ3 serves as a controlled sensitivity analysis showing that even under a modest public tuning setup, instruction tuning can induce measurable and non-monotonic capability shifts across task types. Nevertheless, the broader trend observed in RQ3 is consistent with our endpoint comparison of Qwen2.5-Coder and DeepSeek-Coder families: instruction tuning tends to improve instruction-oriented or guidance-heavy tasks, while often weakening infilling-oriented tasks that require concise local code insertion.

\textbf{External Threats.}
\textit{(1) Task scope.}
Our evaluation focuses mainly on function-level infilling, completion, and instruction-following benchmarks, which are widely used in the CodeLLM literature and allow controlled comparison across model variants. However, existing coding paradigms have evolved towards sophisticated model scaffolding and harness engineering \cite{openhands,opencode} to address repository-level code generation and editing with high-level instructions (\ie \textit{vibe coding}, hereby defined as \textit{Vibe} mode). Notably, our work is distinctive yet relevant to scenarios where developers have to operate on codebases directly in a human-assisted manner. Therefore, the Flow and Command paradigms do not diminish with the advent of Vibe where developers have minimal exposure to the codebases generated by vibe coding.

\textit{(2) Deployment context.}
Our experiments are conducted in offline benchmark settings, whereas real developers interact with coding assistants through iterative workflows involving partial acceptance, editing, feedback, and repeated prompting. Such interactions may either amplify or reduce the observed trade-offs. For example, verbose outputs may be less harmful when users can manually select useful fragments, but more harmful in automatic completion or fill-in-the-middle settings where concise and bounded suggestions are expected. Therefore, while our results reveal systematic model-level behavioral differences, future user studies and deployment-oriented evaluations are needed to understand their practical impact in real development environments.


\section{Related Work}

\paragraph{CodeLLMs and Their Evaluation.}
Large language models have rapidly evolved into specialized CodeLLMs that support a broad range of software engineering activities, including code completion \cite{zhuo2024bigcodebench}, infilling \cite{infilling-bavarian}, classification \cite{li2025titanvul}, explanation \cite{lu2021codexglue}, and repair \cite{swebench}. Early evaluation benchmarks such as HumanEval~\cite{humaneval} and MBPP~\cite{mbpp} established the dominant paradigm of measuring functional correctness on relatively self-contained programming problems. These benchmarks allowed further research on improving CodeLLMs' abilities on various coding tasks \cite{magiccoder,starcoder2}, diversifying CodeLLMs' usage paradigms from simple code generations to interactive pair programming scenarios \cite{barke2023grounded, Muennighoff24octopack}.

\paragraph{Instruction Tuning for CodeLLMs.}
Alongside benchmark development, a separate line of work has investigated how code LLMs should be adapted to better serve natural language-driven programming tasks. OctoPack~\cite{Muennighoff24octopack} demonstrates that instruction tuning can substantially expand the usability of code models by leveraging instruction-like supervision from large-scale commit data, thereby improving performance across diverse coding tasks beyond pure code completion. Similarly, industrial deployments such as CodeCompose show that fine-tuned code models can support real developers at scale across multiple programming languages and authoring surfaces~\cite{murali2024aiassisted}. These advances have helped establish instruction tuning as a standard step in the development of practical code assistants. However, prior work mainly emphasizes the capabilities unlocked by instruction tuning, rather than the trade-offs it may introduce for other coding behaviors. As a result, base and instruct variants are often treated simply as different points on a leaderboard, with limited attention to what is gained, what is lost, and why these changes occur.

\paragraph{Human-Centered Perspectives on AI-Assisted Coding.}
Human-centered studies further suggest that these trade-offs matter because programmers do not interact with code models in a single uniform way. \textit{Grounded Copilot} shows that programmers' interactions with code-generating models are \emph{bimodal}: in \emph{acceleration mode}, users already know what to do and rely on the model to complete or refine code faster, whereas in \emph{exploration mode}, users are uncertain and use the model to propose options, strategies, or higher-level guidance~\cite{barke2023grounded}. Mixed-method evidence from real-world deployments similarly suggests that developers value coding assistants both for speeding up routine authoring and for helping discover APIs, boilerplate patterns, and implementation directions~\cite{murali2024aiassisted}. This perspective is important because these modes correspond to different technical demands: acceleration-oriented use is closely tied to localized continuation and infilling, while exploration-oriented use is more aligned with instruction following and natural-language-mediated assistance. Our work is motivated by precisely this tension. Rather than asking only which model ranks higher overall, we ask how instruction tuning reshapes a code model's behavior across these different coding scenarios. By formalizing the \emph{instruction-tuning tax}, we shift attention from end-point leaderboard comparison to the development trade-offs that arise when adapting code LLMs for assistant-style use.

\section{Conclusion}
In this paper, we conduct the first empirical study that uncovers a key trade-off caused by instruction tuning across programming modes, which we term the \textit{Instruction-Tuning Tax}. Our results show that \textit{instruction tuning is not a free lunch}: teaching a base code model to follow coding instructions improves performance on instruction-oriented tasks, but can also weaken infilling ability and alter generation behavior in flow-mode settings. To better understand this trade-off, we extend our study with both qualitative and quantitative analyses, including manual failure categorization and behavioral metrics such as Code-to-Token Ratio (CTR), natural-language rate, and extra definition/class generation. Our findings show that the instruction-tuning tax is not uniform across model families, with Qwen2.5-Coder generally remaining more robust than DeepSeek-Coder on infilling and completion tasks. Although instruction-tuned models tend to benefit more from structured natural-language guidance, base models still retain clear advantages on fill-in-the-middle tasks. Furthermore, intermediate-checkpoint analysis shows that instruction tuning leads to non-monotonic yet systematic capability reallocation across tasks: instruction-following ability improves, infilling tends to degrade, and completion remains positive on aggregate. Overall, our study highlights the need to view instruction tuning not as a universally beneficial upgrade, but as a trade-off that must be carefully considered when developing and deploying AI coding assistants.

\section{Data Availability}
The artifacts for this study are available at \href{https://github.com/arkosioscambions/CodeTalkers}{https://github.com/arkosioscambions/CodeTalkers} for replication and future research.
\balance
\bibliographystyle{ACM-Reference-Format}
\bibliography{acmart}

\end{document}